\documentclass[twocolumn,showpacs,amssymb,prd,superscriptaddress,nofootinbib]{revtex4-1}
\usepackage{graphicx, epsfig, amssymb} 
\usepackage{amsmath, amsfonts}
\usepackage{bm} 

\usepackage[linktocpage]{hyperref}
\usepackage[caption=false]{subfig}

\def\be{\begin{equation}}
\def\ee{\end{equation}}
\def\beq{\begin{eqnarray}}
\def\eeq{\end{eqnarray}}
\def\non{\nonumber}

\def\p{\partial}

\renewcommand{\vec}[1]{\boldsymbol{#1}}

\begin{document}

\title{Black holes in a box: \\Towards the numerical evolution of black holes in AdS 
       space-times}

\author{Helvi~Witek}\email{helvi.witek@ist.utl.pt}
\affiliation{Centro Multidisciplinar de Astrof\'{\i}sica --- CENTRA, Departamento de F\'{\i}sica, Instituto Superior T\'ecnico, Universidade T\'ecnica de Lisboa - UTL, Av. Rovisco Pais 1, 1049-001 Lisboa, Portugal}

\author{Vitor~Cardoso}\email{vitor.cardoso@ist.utl.pt}
\affiliation{Centro Multidisciplinar de Astrof\'{\i}sica --- CENTRA, Deptartamento de F\'{\i}sica, Instituto Superior T\'ecnico, Universidade T\'ecnica de Lisboa - UTL, Av. Rovisco Pais 1, 1049-001 Lisboa, Portugal}
 
\affiliation{Department of Physics and Astronomy, The University of Mississippi, University, MS 38677-1848, USA}

\author{Carlos Herdeiro}\email{herdeiro@ua.pt}
\affiliation{Departamento de F\'\i sica da Universidade de Aveiro, 
Campus de Santiago, 3810-183 Aveiro, Portugal.}
\affiliation{Centro de F\'\i sica do Porto --- CFP,
Departamento de F\'\i sica e Astronomia,
Faculdade de Ci\^encias da Universidade do Porto --- FCUP,
Rua do Campo Alegre, 4169-007 Porto, Portugal.}

\author{Andrea~Nerozzi}\email{andrea.nerozzi@ist.utl.pt}
\affiliation{Centro Multidisciplinar de Astrof\'{\i}sica --- CENTRA, Dept.\ de F\'{\i}sica, Instituto Superior T\'ecnico, Universidade T\'ecnica de Lisboa - UTL, Av. Rovisco Pais 1, 1049-001 Lisboa, Portugal}

\author{Ulrich~Sperhake}\email{sperhake@tapir.caltech.edu}
\affiliation{Department of Physics and Astronomy, The University of Mississippi, University, MS 38677-1848, USA}
\affiliation{Institut de Ci\`encies de l'Espai (CSIC-IEEC), Facultat de Ci\`encies, Campus UAB, E-08193 Bellaterra, Spain}

\affiliation{California Institute of Technology, Pasadena, CA 91125, USA}

\author{Miguel~Zilh\~ao}\email{mzilhao@fc.up.pt}
\affiliation{
  Centro de F\'\i sica do Porto --- CFP, 
  Departamento de F\'\i sica e Astronomia, 
  Faculdade de Ci\^encias da Universidade do Porto --- FCUP, 
  Rua do Campo Alegre, 4169-007 Porto, Portugal
}

\date{\today}

\begin{abstract}
  The evolution of black holes in ``confining boxes'' is interesting for a
  number of reasons, particularly because it mimics the global structure of
  Anti-de Sitter geometries. These are nonglobally hyperbolic space-times and
  the Cauchy problem may only be well defined if the initial data is
  supplemented by boundary conditions at the timelike conformal boundary. Here,
  we explore the active role that boundary conditions play in the evolution of a
  bulk black hole system, by imprisoning a black hole binary in a box with
  mirrorlike boundary conditions.
  We are able to follow the post-merger dynamics for up to two reflections off
  the boundary of the gravitational radiation produced in the merger. We
  estimate that about 
  15\% of the radiation energy is absorbed by the black
  hole per interaction, whereas transfer of angular momentum from the radiation
  to the black hole is only observed in the first interaction. We discuss the
  possible role of superradiant scattering for this result. Unlike the studies
  with outgoing boundary conditions, both the Newman-Penrose scalars $\Psi_4$
  and $\Psi_0$ are nontrivial in our setup, and we show that the numerical data
  verifies the expected relations between them.
\end{abstract}



\maketitle


\clearpage
\section{Introduction}
Numerical relativity and the gauge/gravity correspondence are two fields
in high energy/gravitational physics that have seen tremendous activity
and progress over the last few years. Since the 2005 breakthroughs
\cite{Pretorius:2005gq,Campanelli:2005dd,Baker:2005vv}
the numerical relativity community has produced stable evolutions of black
hole (BH) binaries in 4-dimensional, asymptotically flat space-times,
covering the inspiral, merger and ring-down phases. State of the art
simulations can be as long as 15 orbits plus merger and ring-down
\cite{Boyle2007},
can deal with large mass ratios \cite{Gonzalez2008,Lousto:2010tb,Lousto:2010qx,
Lousto:2010ut},
with dimensionless BH spins
of up to 0.92 \cite{Dain:2008ck}
or with eccentricities as low as $5\times 10^{-5}$
(see \cite{Hinder:2010vn} for a recent review). Such simulations have
unveiled new phenomena such as BH \textit{kicks}
\cite{Gonzalez:2006md,Gonzalez2007a,Campanelli:2007cga}
and some groups are
now focusing on the construction of sufficiently accurate template
waveforms to be used in filtering data from the current (LIGO, Virgo,
GEO, TAMA) and planned (Advanced LIGO and LISA) gravitational wave
observatories \cite{Ajith2007, Ajith2007a,Buonanno2007a,Buonanno2009}.
Simultaneously, high energy collisions of BHs have
been simulated \cite{Sperhake:2008ga,Sperhake:2009jz,Shibata:2008rq},
leading to accurate results for the scattering cross section and for the
conversion rate of the initial centre-of-mass energy into gravitational
radiation. These simulations have also tested cosmic censorship, the
Dyson luminosity limit  and exhibited the zoom-whirl behaviour first
found in \cite{Pretorius:2007jn}.

The gauge/gravity correspondence has been developed, since the original
proposal by Maldacena in 1998 \cite{Maldacena:1997re}, both as an
efficient tool to understand strongly coupled gauge theories using
classical gravity and as way to study gravitational phenomena from a
dual field theory. Central to many of these developments are BHs, as may
be seen by the following list of examples: (i) the successful
microscopic computations of the Bekenstein-Hawking entropy for extremal
BHs \cite{Strominger:1996sh}
and Hawking emission rates for near extremal BHs \cite{Callan:1996dv}
are now seen as applications of the correspondence; (ii) the
confinement/deconfinement phase transition in QCD-like theories has been
identified \cite{Witten:1998qj} with the Hawking-Page phase transition for
Anti-de-Sitter (AdS) BHs \cite{Hawking:1982dh}; (iii)  moving
away from thermal equilibrium, the quasinormal frequencies of AdS BHs
have been identified with the poles of retarded correlators describing
the relaxation back to equilibrium of a perturbed dual field theory
\cite{Horowitz:1999jd,Birmingham:2001pj}; (iv) in a large class of
gauge theories with a gravity dual, a universal behaviour was obtained for
the ratio of the strongly coupled medium's viscosity to entropy density,
by computing the absorption cross-section of low energy gravitons in
the dual BH (or black brane) geometry \cite{Kovtun:2004de}. The result
is in good agreement with experimental results from the Relativistic
Heavy Ion Collider (RHIC); (v) critical exponents, of the type
found in spherical gravitational collapse by Choptuik in 4-dimensional
asymptotically flat space-time \cite{Choptuik:1992jv}, have been
conjectured to be dual to the asymptotic value of the parton saturation
exponent for high energy scattering in QCD, in the Regge limit, at weak
coupling \cite{AlvarezGaume:2006dw}. These and other examples have built
expectations that the gauge/gravity correspondence will be a useful 
laboratory for gaining understanding of very difficult problems of both 
field theory and gravity, such as the confinement problem, 
the information loss paradox and the problem of singularities.

Given the potential of the correspondence exemplified above, a working
framework to solve Einstein's equations exactly for a broad range
of initial conditions in AdS spaces would, unquestionably, be very
useful for deepening the study of the correspondence, particularly in more
dynamical situations. To assemble such a framework is our long-term goal
and the present paper serves as the first step in this direction,
i.e., to generalize the techniques of numerical relativity to AdS
space-times. In order to achieve this goal, one has to go beyond the 
standard methods of numerical relativity in, at least, two obvious points.

Firstly, AdS space-times are not globally hyperbolic. In asymptotically
AdS spaces the boundary plays an  ``active role'' for the bulk
evolution. This is easily visualised in the Penrose diagram of
AdS, which has a timelike boundary. Physically, null geodesics
in AdS reach the boundary for a finite affine parameter. One
thus often refers to an asymptotically AdS space as a ``box'',
having in mind that AdS boundary conditions directly affect the
bulk physics \cite{Wald:1980jn,Ishibashi:2003jd,Ishibashi:2004wx}.
This should be contrasted with the asymptotically flat case, where
the only physically relevant choice for the boundary conditions
of the bulk fluctuations corresponds to outgoing waves at spatial
infinity.  In the gauge/gravity correspondence, the choice of the
AdS boundary conditions is dictated by a holographic prescription
\cite{Son:2002sd,Herzog:2002pc,Skenderis:2008dh,Skenderis:2008dg}.
Secondly, from the viewpoint of the duality, $D$-dimensional AdS
space-times, AdS$_D$  (not just AdS$_4$) are relevant. Thus, we would
like to have a framework that could be used in AdS$_D$, in particular
for $D=5$, which is related by the correspondence to 4-dimensional gauge
theories. The latter issue has been recently addressed by our group in
separate publications \cite{Zilhao:2010sr,Witek:2010xi}, as well as by other groups
\cite{Yoshino:2009xp,Sorkin:2009wh,Sorkin:2009bc,Nakao:2009dc,Shibata:2009ad,Lehner:2010pn}. 
Here we shall focus
on the former issue: the active role of boundary conditions.

The dynamics of BHs in AdS, and especially the role of spatial infinity
(``the box'') is poorly understood. In contrast to the asymptotically
flat case, interesting new phenomena may occur in AdS backgrounds.
For instance, superradiance effects have been shown to make small (as
measured by the AdS radius)
rotating BHs unstable, through a sequence of
reflections at the boundary and amplifications close to the ergoregion
\cite{Hawking:1999dp,Cardoso:2004hs,Cardoso:2004nk,Cardoso:2006wa,
Kodama:2009rq,Murata:2008xr,Kodama:2007sf,Aliev:2008yk,Uchikata:2009zz}.
The final state of this instability could be a new nonaxisymmetric BH
configuration, which is also supported by recent gravity/hydrodynamics
arguments \cite{Cardoso:2009bv,Cardoso:2009nz}. Notice that
nonaxisymmetric BHs are strictly forbidden in asymptotically flat
space-times \cite{Hollands:2006rj,Hollands:2008wn}, so the boundary does
have an important role in the description of BHs.

In order to identify in the cleanest possible way the active role of
the boundary for the bulk evolution, we consider here a toy model for
AdS. We set the cosmological constant to zero and impose mirrorlike
boundary conditions on a box that contains the dynamical system. This
mimics the AdS global geometry, keeping the local geometry of vacuum
models. We choose the dynamical system to be a BH binary, starting
at some given distance, producing either a head-on collision or an
inspiralling merger. In the latter situation we consider the initial
BHs without intrinsic angular momentum. These are, by now,
very well tested systems when purely outgoing boundary conditions are
imposed. Thus we will be able to see clearly the modifications due to
the nonoutgoing boundary conditions in systems with nontrivial dynamics.

For the post-merger dynamics the inspiralling binaries provide a more
interesting analysis than the head-on collision case, since the initial
centre-of-mass energy transferred into gravitational radiation is by more 
than one order of magnitude larger in the former case as compared to
the latter. Immediately after the merger, the system will contain a 
single (spinning or nonspinning) BH plus gravitational
radiation. This radiation will then be (repeatedly) reflected off the
boundary and interact  with the BH.

The first nontrivial result is that we \textit{can} follow the
numerical evolution for up to two reflections off the boundary of the
gravitational radiation produced in the merger. \textit{A priori} it 
was not guaranteed that this could be achieved, since it is not known
whether the formulation of the Einstein equations that we use provides a
well defined initial value boundary problem together with the boundary
conditions we impose. Our simple setting actually provides a first attempt
to test the well-posedness of the initial boundary value problem in a
non globally hyperbolic space-time.  We find that our numerical results
are at least second-order convergent for at least two  reflections off
the wall,  after which we gradually lose convergence.
A deeper study of these issues is clearly needed, as
well as an exploration of how the convergence (and remaining results)
change for different boundary conditions.

During the window of numerical convergence, we study the properties of both
outgoing \textit{and ingoing} gravitational radiation. The usual studies of BH
binaries with outgoing boundary conditions, focus only on the Weyl scalar
$\Psi_4$, which describes outgoing gravitational waves. However, an equally
relevant quantity for the description of gravitational radiation is the scalar
$\Psi_0$, which describes ingoing waves, but which is seldom discussed in the
literature. 
Due to our special boundary conditions and setup, we are able to verify 
certain relations between these two quantities in a numerical evolution 
for the first time. This also provides a test on the correctness and meaning 
of the boundary conditions we have imposed. 

By analysing the properties of the apparent horizon of the BH produced in the
merger and after each interaction with the gravitational wave packet, we
estimate the amount of energy and angular momentum that is transferred from the
radiation into the BH per interaction. 
In case of the inspiralling binary the boxed BH is spinning and we expect
superradiant scattering of the waves generated during merger to become
important; in fact, the back and forth bouncing of the waves at the reflecting
wall and their subsequent amplification by superradiance close to the ergoregion
are expected to turn the system into a BH
bomb~\cite{Press:1972,Cardoso:2004hs,Cardoso:2004nk}\footnote{The artificial
  mirror sometimes appears naturally. A massive scalar field scattering off a
  Kerr BH acts as its own reflecting wall
  \cite{Damour:1976,Detweiler:1980uk,Zouros:1979iw,Furuhashi:2004jk,
    Strafuss:2004qc,Hod:2009cp,Rosa:2009ei}.  Furthermore, Kaluza-Klein modes in
  dimensional reduction can also act as an effective mass rendering
  higher-dimensional, rotating BHs unstable
  \cite{Cardoso:2004zz,Cardoso:2005vk}. Finally, it has been suggested that
  astrophysical BHs might sometimes behave as BH bombs, with the role of the
  reflecting cavity being played by accretion disks
  \cite{Putten:1999,Aguirre:1999zn}.}. Thus, these simulations will be the
first attempt at a nonlinear study of the BH bomb. An important open problem is
understanding how the evolution proceeds and what is the end point of the
instability. This can only be achieved through nonlinear studies. A final
statement on this issue will, however, require further analysis than that
provided herein.


This paper is organized as follows. Section~\ref{sec:formulation} briefly
reviews the setup to evolve Einstein's equations numerically for the case at
hand, including a brief description of the numerical code, of the formulation of
Einstein's equations in a so-called BSSN form, of the gauge choice adopted, the
way the ``spherical'' boundary is imposed and the boundary conditions. In
Section~\ref{sec:extraction} we explain which of the numerical outputs we use to
extract relevant physical quantities, in particular gravitational wave estimates
(Section~\ref{sec:waveextraction}) and apparent horizon (AH) estimates
(Section~\ref{sec:AH}). The numerical results of our simulations are shown in
Section~\ref{sec:num}.  In Section~\ref{sec:conclusions} we close with some
discussion of the results and prospects for the future. Some technical
points and further results have been organised into three appendices. For self-containedness,
Appendix~\ref{sec:appWaveExtraction} reviews the electromagnetic decomposition
of the Weyl tensor and in particular the construction of the relevant quantities
for our study, $\Psi_0$ and $\Psi_4$. Appendix~\ref{sec:appsnapshots} exhibits some snapshots for
visualising the evolution of the system we have studied. Appendix~\ref{sec:appcubebox} describes the simulations with a cubic,
rather than spherical, box. 
 
\section{Numerical framework\label{sec:formulation}}
In order to numerically generate a solution to the Einstein field equations,
it is most convenient to view the problem as a time evolution or 
{\em initial value problem}. The majority of formulations of the
Einstein equations as an evolution system in time is based on
the canonical ``3+1''-decomposition introduced by Arnowitt, Deser and Misner
(ADM) \cite{Arnowitt:1962hi} and further developed by York \cite{York1979}.
One thus obtains a first order, constrained
evolution system in time for six components
each of the three-metric $\gamma_{ij}$ and the extrinsic curvature
$K_{ij}$ which describe the intrinsic geometry of three-dimensional
hypersurfaces as well as their embedding in the four-dimensional
space-time. The {\em Hamiltonian} and {\em momentum} constraints
impose four conditions on $\gamma_{ij}$ and $K_{ij}$ on each
hypersurface but are conserved under the time evolution. Finally,
four gauge variables, the lapse $\alpha$ and the shift $\beta^i$
represent the coordinate freedom of Einstein's relativity. Suitable
specification of these free variables is crucial for a successful
numerical implementation.

Our numerical framework is based on a method now commonly
referred to as {\em moving punctures} \cite{Campanelli:2005dd,
Baker:2005vv}. The Einstein equations are formulated
as the Baumgarte-Shapiro-Shibata-Nakamura (BSSN)
system \cite{Shibata:1995we, Baumgarte:1998te}, a modification
of the ADM formulation which employs the variables
\begin{eqnarray}
  \chi &=& \psi^{-4} = \gamma^{-\frac{1}{3}}\,, \,\,\,
           \tilde{\gamma}_{ij} = \chi \gamma_{ij}\,,\non\\
     K &=& \gamma^{ij} K_{ij}\,,\,\,\,
           \tilde{A}_{ij} = \chi A_{ij} = \chi \left(K_{ij}
           - \frac{1}{3}\gamma_{ij}K \right)\,,\non\\
 \tilde{\Gamma}^i &=& \tilde{\gamma}^{jk}\tilde{\Gamma}_{jk}^i
           = -\p_j \tilde{\gamma}^{ij}\,.
  \label{eq:BSSNvars}
\end{eqnarray}
By construction $\det \tilde{\gamma}_{ij}=1$ which implies
the last equality for $\tilde{\Gamma}^i$. The exact form
of the evolution equations for this set of variables
is given in Eqs.~(A1, A4, A6, A7, A8)\footnote{Note that the
final term on the right hand side of their Eq.~(A6) should
be $\frac{2}{3}\chi (\alpha K - \partial_m \beta^m)$, i.~e.~a
factor of $\chi$ is missing.} of Ref.~\cite{Sperhake:2006cy}.
Finally, we evolve the gauge variables $\alpha$ and $\beta^i$
using ``1+log'' slicing and a $\Gamma$-driver of the form
\begin{eqnarray}
  \partial_t \alpha &=& \beta^m \partial_m \alpha - 2\alpha K, \\
  \partial_t \beta^i &=& \chi_{\beta} B^i, \\
  \partial_t B^i &=& \partial_t \tilde{\Gamma}^i - \eta_{\beta} B^i.
\end{eqnarray}
Here $\chi_{\beta}$ and $\eta_{\beta}$ are constant parameters set to
one for all simulations reported in this work.

We evolve these equations with the {\sc Lean} code \cite{Sperhake:2006cy} which
is based on the \textsc{Cactus} computational toolkit~\cite{cactus} and the
\textsc{Carpet} mesh refinement package \cite{Schnetter:2003rb,
  carpet}. BH binary initial data are provided by the spectral solver of
Ansorg {\em et al.}  \cite{Ansorg:2004ds} and the calculation of apparent
horizons is performed with Thornburg's {\sc AHFinderDirect}
\cite{Thornburg:1995cp, Thornburg:2003sf}. For more details of the code we refer
the reader to Ref.~\cite{Sperhake:2006cy} and Sec.~III of
\cite{Sperhake:2007gu}.

The key ingredient in which our current numerical framework differs from
previous implementations of the {\sc Lean} code and most other codes is
the outer boundary condition, which we will discuss in more detail in the
remainder of this section.

The vast majority of numerical simulations
of BH binaries has been concerned with asymptotically flat space-times
and consequently employed either of the following boundary treatments:
(i) outgoing Sommerfeld conditions on Cartesian grids of finite size,
as described for example in \cite{Alcubierre:2002kk}; (ii) outgoing radiation
\cite{Pollney:2009yz, Pollney:2009ut} with multipatch methods,
including {\em Cauchy characteristic} wave extraction
\cite{Reisswig:2009us,Reisswig:2009rx} and (iii)
constraint preserving boundary conditions combined with multidomain
methods~\cite{Rinne:2007ui,Pazos:2009vb}.

\begin{figure}
\includegraphics[width=0.5\textwidth]{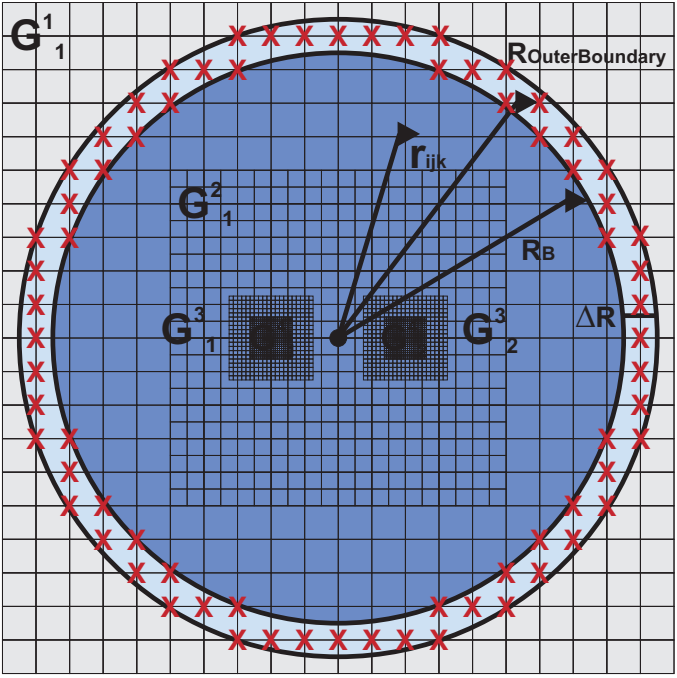}
\caption{\label{fig:SphBD} Illustration of a (Lego-)spherical
outer boundary.}
\end{figure}

\begin{figure}
\includegraphics[width=8cm]{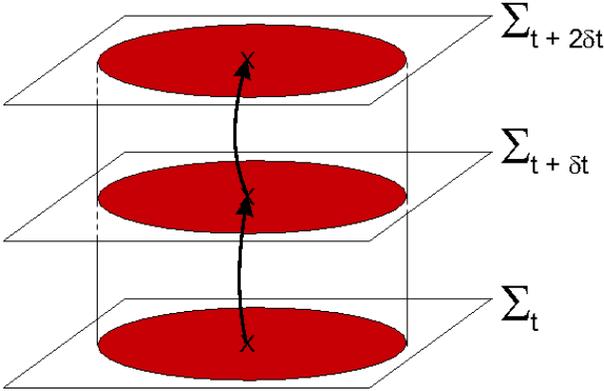}
\caption{\label{fig:foliation} Sketch of the foliation for the numerical
evolution of BH binaries in a (spherical)  box.  The location of the
considered numerical domain on each spatial hypersurface is shown as a
dark (red) sphere.
}
\end{figure}

In contrast we will study the dynamics of BH space-times under
the influence of a reflective outer boundary. It is natural to use
for this purpose an outer boundary of spherical shape. Most importantly,
this avoids mixing of different gravitational wave multipoles as would
occur in the case of a reflective, cubic outer boundary. This is
discussed in more detail in Appendix~\ref{sec:appcubebox} where we
compare simulations using both types of boundary. Except for this
comparison, however, we will exclusively study spherical
outer boundaries or, rather, approximate these by using so-called {\em
Lego} spheres; cf.~Sec.~3 in \cite{Shoemaker:2003td}.
In Fig.~\ref{fig:foliation} we sketch the foliation of the space-time
under consideration, suppressing one spatial dimension for simplicity. 
The numerical domaine, i.e., the Lego sphere is visualized by a dark (red) 
domain on each timeslice $\Sigma_{t+n\delta t}$. 
Their numerical implementation is illustrated in Fig.~\ref{fig:SphBD}
which schematically displays a computational
domain using four refinement levels with one or two components each. The
individual components are labelled $G^i_m$ where the indices $i$
and $m$ denote the refinement level and component number.
Note that one spatial dimension is suppressed for visualisation purposes.
In order to update a grid function at a particular vertex, we
require information from neighbouring points because of the discretization
of spatial derivatives
in the evolution equations. The exact number $n$ of neighbouring points
required in each direction depends on the finite difference stencils
employed. While $n=3$ for the $4^{\rm th}$ order accurate stencils
used in our simulations, we use $n=1$ for simplicity in our illustration in
Fig.~\ref{fig:SphBD}. Consider first the dark (blue) shaded area inside the
inner solid circle of radius $R_B$. Each point in this {\em regular}
domain can be updated straightforwardly
provided we also have valid data on the boundary points marked by $\times$
symbols. Points outside the 
circle of radius $R_B$ are not
required for updating regular points and are simply ignored in the
numerical evolution. 
The specific boundary condition is then determined
by the manner in which we update grid functions on the boundary points
marked as $\times$ in the figure.

In order to mimic the global structure of an Anti-de Sitter space-time
we effectively enclose the BH binary inside a spherical mirror
and set
\begin{equation}
\label{eq:refbc}
\frac{\partial}{\partial t} f = 0,
\end{equation}
at each boundary point with $f$ denoting any of the BSSN variables
listed in Eq.~(\ref{eq:BSSNvars}).
The use of fourth-order stencils adds one complication to this picture:
the upgrade of a grid point requires two neighbors, so that points
right next to the boundary need special treatment. In practice, we have
achieved optimal stability properties by evolving these points with
second-order stencils.
%
Our implementation requires one further ingredient in order
to handle the spurious radiation inherent to numerically generated
initial data of BH binary systems; cf.~\cite{Bode:2007dv}.
In order to avoid contamination of our simulations by such spurious
radiation being trapped inside our reflective boundary we employ
standard outgoing radiation boundary conditions at early times
and only switch on our reflective condition at
\begin{equation}
  t_{\rm ref} = R_B +\Delta t_{\rm pulse}.
\end{equation}
%
%
In order to avoid a discontinuous jump from outgoing to reflective
boundary conditions, we gradually switch off the time derivative
$\partial f / \partial t$ using a weighting factor $w(t)$ which smoothly
decreases from 1 to 0 over an interval 
$\Delta t = 10~M$ and $\Delta t = 20~M$ for the head-on collision and inspiral, respectively.
The duration of the spurious wave pulse $\Delta t_{\rm pulse}$
is estimated from previous simulations of similar setups in
asymptotically flat space-times as for example presented in
Refs.~\cite{Baker:2006yw,Sperhake:2006cy,Berti:2007fi}.
The spurious radiation is thus given sufficient time to leave
the computational domain.
%
\section{Extraction of meaningful quantities}\label{sec:extraction}
\subsection{Wave extraction}\label{sec:waveextraction}
Information about gravitational radiation is most commonly obtained in
numerical simulations of BH binaries by computing the Weyl
scalar $\Psi_4$, which has the asymptotic property of being equal to
the outgoing radiation if the complex null-tetrad is chosen properly. In a
truly asymptotically AdS space-time this procedure cannot be implemented
so straightforwardly. In our toy model, which has the same local geometry
as vacuum models, for a sufficiently large box size we expect the standard
tetrad to yield the correct gravitational wave information. This is
yet another advantage of our setup. Moreover, because we also deal with
ingoing waves in our simulations, once they are reflected
from the box boundary, we will consider the Weyl scalar $\Psi_0$ as well,
in order to account for the ingoing contribution.
The authors are not aware of any numerical study of the ingoing Weyl
scalar $\Psi_0$.
We will therefore analyse it in detail in the course of our numerical studies.

To be explicit, we define a spherical coordinate system centred on
the centre of mass of the binary with orthonormal basis
$(\hat{r},\hat\theta,\hat\phi)$.
The coordinates are chosen such that the azimuthal axis is aligned with
the orbital angular momentum
and the binary orbits are in the direction of increasing azimuthal coordinate. 
Our definitions and notation are the same as in
\cite{Friedrich:1996hq,Sperhake:2006cy}.
To define our complex null-tetrad, we use the timelike unit vector normal
to a given hypersurface $\hat{n}$ and the radial unit vector $\hat{r}$
to define an ingoing ($\vec{k}$) and outgoing null vector ($\vec{\ell}$) by
\be
\label{eq:nulltetrad1}
\vec{k} \equiv \frac1{\sqrt{2}}(\hat{n} + \hat{r})\,,\quad
\vec{\ell} \equiv \frac1{\sqrt{2}}(\hat{n} - \hat{r})\,.
\ee
We define the complex null vector $\vec{m}$ and its complex conjugate by
\be
\label{eq:nulltetrad2}
\vec{m} \equiv \frac1{\sqrt{2}}(\hat\phi + i\hat\theta), \quad 
\vec{\bar{m}} \equiv \frac1{\sqrt{2}}(\hat\phi - i\hat\theta).
\ee
In terms of this tetrad, we define $\Psi_0$ and $\Psi_4$ as
\beq
\label{eq:Psi0_Weyl}
  \Psi_0 &\equiv & C_{\alpha\beta\gamma\delta}k^\alpha m^{\beta}k^\gamma m^\delta, \\
\label{eq:Psi4_Weyl}
    \Psi_4 &\equiv & C_{\alpha\beta\gamma\delta}\ell^\alpha \bar{m}^{\beta}\ell^\gamma \bar{m}^\delta,
\eeq
where $C_{\alpha\beta\gamma\delta}$ is the Weyl tensor.
To relate $\Psi_0$ and $\Psi_4$ to the amplitudes of the gravitational
waves, we note that in the transverse-traceless (TT) gauge, assuming the
functional form $f(t \pm r)$ for the ingoing or outgoing waves, we have
\beq
\frac14(\ddot{h}^{TT}_{\hat\theta\hat\theta}-\ddot{h}^{TT}_{\hat\phi\hat\phi}) &=&
-R_{\hat{n}\hat\theta\hat{n}\hat\theta} =\mp R_{\hat{n}\hat\phi\hat{r}\hat\phi}=
-R_{\hat{r}\hat\theta\hat{r}\hat\theta} \nonumber \\
&=&R_{\hat{n}\hat\phi\hat{n}\hat\phi}=\pm R_{\hat{n}\hat\theta\hat{r}\hat\theta} =
R_{\hat{r}\hat\phi\hat{r}\hat\phi}\,, \nonumber
\eeq
\beq
\frac12\ddot{h}^{TT}_{\hat\theta\hat\phi} &=&
-R_{\hat{n}\hat\theta\hat{n}\hat\phi} =
-R_{\hat{r}\hat\theta\hat{r}\hat\phi} =
\pm R_{\hat{n}\hat\theta\hat{r}\hat\phi} =
\pm R_{\hat{r}\hat\theta\hat{n}\hat\phi}\nonumber\,.
\eeq
A `dot' denotes derivative with respect to the argument.  Following
standard conventions, we take the $h_+$ and $h_\times$ polarisations
of the gravitational waves to be given by
\be \ddot{h}_+ = \frac12(\ddot{h}^{TT}_{\hat\theta\hat\theta} 
- \ddot{h}^{TT}_{\hat\phi\hat\phi})\,,\quad
\ddot{h}_\times = \ddot{h}^{TT}_{\hat\theta\hat\phi}\,.
\ee
Then, we find that in vacuum regions of the space-time, for outgoing waves
$ \ddot{h}_+= \ddot{h}_+(t-r)$ and $\ddot{h}_\times=\ddot{h}_\times(t-r)$,
\beq \label{eq:Psi4_ddh_defn} \Psi_0 &=&0\,, \\
\Psi_4 &=& \ddot{h}_+ + i\ddot{h}_\times\,,
\eeq
while for ingoing waves $ \ddot{h}_+= \ddot{h}_+(t+r)$ and $\ddot{h}_\times=\ddot{h}_\times(t+r)$
\beq
\label{eq:Psi4_ddh_defn2} \Psi_0 &=& \ddot{h}_+ - i\ddot{h}_\times, \\
\Psi_4 &=& 0\,.
\eeq
The fact that $\Psi_4$ ($\Psi_0$) are zero for ingoing (outgoing) waves is
consistent with the leading order in perturbation theory. The correct
expression and
in particular its dependence on the radial coordinate is given by the
solution of the
Teukolsky master equation at large distances from the source, which
states that for
outgoing waves \cite{Teukolsky:1973ha}
\be \label{asymptotics}
\Psi_0 \approx \frac{e^{i\omega(t-r)}}{r^5}\,,\quad
\Psi_4 \approx \frac{e^{i\omega(t-r)}}{r}\,,
\ee
while for ingoing waves
\be
\Psi_0 \approx \frac{e^{i\omega(t+r)}}{r}\,,\quad
\Psi_4 \approx \frac{e^{i\omega(t+r)}}{r^5}\,.
\ee
We decompose the resulting $\Psi_4$ ($\Psi_0$) into modes
by projection onto spherical harmonics of spin-weight $s=-2$ ($s=2$)
according to
\beq
\label{psi4dec} Mr\Psi_4&=&Mr\,\sum_{l=2}^\infty \sum_{m=-l}^l
 \,{_{-2}}Y_{lm}(\theta\,,\phi)\, \psi^4_{lm}\,, \\
Mr\Psi_0&=&Mr\,\sum_{l=2}^\infty \sum_{m=-l}^l \,{_{2}}Y_{lm}(\theta\,,\phi)\, \psi^0_{lm}\,,
\eeq
where $_{-2}Y_{l m}(\theta,\phi)$ and $_{2}Y_{l m}(\theta,\phi)$
are spin-weight $-2$
and $2$ spherical harmonics \cite{Goldberg:1966uu,Berti:2005gp}.
These are defined as
\be
_sY_{l m}(\theta,\phi) \equiv
(-1)^s\sqrt{\frac{2l+1}{4\pi}}d^l_{m(-s)}(\theta)e^{im\phi},
\ee
where $d^l_{ms}$ is the Wigner $d$-function 
\begin{equation}
\begin{split}
  d^l_{ms}(\theta) & \equiv
  \sum_{t=C_1}^{C_2}\frac{(-1)^t\sqrt{(l+m)!(l-m)!(l+s)!(l-s)!}}
  {(l+m-t)!(l-s-t)!t!(t+s-m)!} \\
  & \quad (\cos\theta/2)^{2l+m-s-2t}(\sin\theta/2)^{2t+s-m}\,,
\end{split}\label{eq:3}
\end{equation}
and where $C_1= \max(0,m-s)$ and $C_2=\min(l+m,l-s)$. 
Here $M$ is the ADM mass of the system, computed from the initial data and assuming this is an asymptotically flat space-time,
and $r$ is the generalised harmonic radial coordinate.



In the numerical code, the null-tetrad is constructed from a Cartesian
orthonormal triad $(u,v,w)$ and the timelike vector $\hat{n}$ is orthonormal to
$t={\rm constant\ }$ hypersurfaces.
The space-time is evolved with time $t$ using Cartesian coordinates $x,y,z$.
In practice, we compute the Newman-Penrose scalars $\Psi_0$ and $\Psi_4$
using the electromagnetic decomposition of the Weyl tensor
according to Eqs.~(\ref{eq:elmagnPsi})
on the entire Cartesian grid.
Then, they are interpolated onto coordinate spheres of different extraction
radii $r_{\rm ex}$ with a uniform distribution of points in 
$(\theta,\phi)$.\footnote{$108\times54$ points in 
$\theta\in [0,\pi]$, $\phi\in [0,2\pi]$ for the set of simulations using
low resolution.
In case of the higher resolutions these numbers are
adjusted accordingly.}
All the waveform related data from the simulations presented
in the course of this paper  are taken from such samplings of
$\Psi_0(t,r=r_{\rm ex},\theta,\phi)$ and $\Psi_4(t,r=r_{\rm ex}, \theta,\phi)$.
A more detailed description is given in Appendix~\ref{sec:appWaveExtraction}.

As discussed previously, $\Psi_4$ is no longer simply related to the energy flux,
but since we are dealing with fairly large box sizes, one might hope
that many notions can be retained in an approximate sense. 
Given the Newman-Penrose scalar $\Psi_4$, we can compute the radiated energy, linear and angular momentum from the radiation
content~\cite{Alcubierre:2008}:
\begin{align}
\label{eq:energywave}
\frac{dE}{dt} & = \lim_{r\rightarrow \infty} \frac{r^2}{16\pi}
      \int_{\Omega} \left| \int_{-\infty}^{t} \Psi_4 d\tilde{t}
      \right|^2 d{\Omega} \,, \\
\frac{dP_i}{dt} & = -\lim_{r \rightarrow \infty} \frac{r^2}{16\pi}
      \int_{\Omega} \ell_i \left| \int_{-\infty}^{t} \Psi_4 d\tilde{t}
      \right|^2 d{\Omega} \,, \\
    \begin{split}
      \frac{dJ_z}{dt} & = -\lim_{r \rightarrow \infty} \frac{r^2}{16\pi}\times \\
      & \quad \mathrm{Re}\left[ \int_{\Omega} \left( 
          \int_{-\infty}^t \Psi_4 d\tilde{t} \right)
          \partial_{\phi} \left( \int_{-\infty}^t \int_{-\infty}^{\hat{t}} \bar{\Psi}_4
          d\tilde{t} d\hat{t}
        \right)d{\Omega} \right] \,
    \end{split}
\end{align}
where
\be
  \ell_i = \left(-\sin \theta \cos \phi,\,-\sin \theta \sin \phi,\,
           -\cos \theta \right)\,.\nonumber
\ee
The definitions above are based on time integrals which start in the infinite
past (at retarded time $t=-\infty$), and thus capture the complete
gravitational wave signal.  Starting the time integrations at $t=-\infty$
corresponds to the limit of infinite extraction radius on the initial time
slice --- the slice would then extend all the way to spatial infinity, no part
of the waveform would be lost, and it would take an infinite time for the
waves to reach the extraction sphere. This situation cannot be handled
with the current numerical
codes; we therefore work with finite extraction radii. 

The mass and angular momentum of the final BH can be estimated from
balance arguments.
Given the parameters $P_{y_{i}},d$ in the Bowen-York initial data, we
straightforwardly calculate the total initial angular momentum as
\be
J_{\rm ini} = L_{\rm ini} = d P_{y_i}\,,
\ee
since the initial spin of each BH is zero.
Ansorg's \textsc{TwoPunctures} \cite{Ansorg:2004ds} initial data solver
directly provides the total ADM mass $M$ of the system and we obtain
radiated energy and angular momentum $E_{\rm rad}$ and $J_{\rm rad}$
from the gravitational wave signal. In case of a merger, this gives
us the final angular momentum and mass of the BH
\beq M_{\rm fin} &=& M - E_{\rm rad}\,, \\
J_{\rm fin} &=& J_{\rm ini} - J_{\rm rad}\,.
\eeq
The dimensionless spin parameter of the final hole follows directly from
\be
j_{\rm fin} = \frac{J_{\rm fin}}{M_{\rm fin}^2}.
\ee
We check our results by fitting the quasinormal frequency and damping
time of the
final BH and invert them to obtain $j_{\rm QNM}$ (see
e.g.~\cite{Berti:2005ys, Berti:2007dg,Berti:2009kk}).

\subsection{Apparent horizon properties}\label{sec:AH}
We can also characterise the process by the properties of the apparent
horizon of the final BH itself. Since this relies only on local
quantities, it does not depend upon the space-time being asymptotically
flat. In order to monitor the mass and spin of the final BH, we use
Thornburg's Apparent
Horizon Finder \textsc{AHFinderDirect}
\cite{Thornburg:2003sf,Thornburg:1995cp} in different ways, which also
allow us to obtain uncertainty estimates:
\begin{enumerate}
\item
  The irreducible mass $M_{\rm irr}$ enables us to calculate the final BH mass $M_{\rm BH}$
  from Christodoulou's relation~\cite{Christodoulou:1970wf}
  \begin{equation}
     M_{\rm BH}^2 = M_{{\rm irr}}^2 + \frac{J^2}{4M_{{\rm irr}}^2}.
             \label{eq:christodoulou}
  \end{equation}

  This relation provides a method to check the internal
  consistency of the result for the final BH spin as calculated from the above
  balance arguments. For this purpose we set $M_{\rm BH}=M_{\rm fin}$ and solve
  Eq.~(\ref{eq:christodoulou}) for the spin
  \begin{equation}
    j_{\rm fin}^2 = \frac{J^2}{M_{\rm fin}^2}
         = 4\frac{M_{\rm irr}^2}{M_{\rm fin}^2}\left(1
           - \frac{M_{\rm irr}^2}{M_{\rm fin}^2}\right)\,,
  \end{equation}

  For comparison we also compute the spin of the final hole from the two
  following estimates:

  \item
    We measure the ratio $C_r(j) = C_p/C_e$ of polar to
  equatorial circumference of the final BH \cite{Anninos:1994pa}.
  If we assume the final
  object to be a Kerr BH, this ratio is
  $C_r=\frac{2}{\pi}\sqrt{1-\beta^2}\,E(\beta^2)$, where
  $\beta^2\equiv\,j^2M/(2r_+)$, $E(\beta^2)$ is a complete elliptic
  integral and $r_+/M=1+\sqrt{1-j^2}$. This expression can be inverted
  to find the dimensionless spin parameter, $j_{C_r}$, of the final hole.

\item
  The equatorial circumference of a Kerr BH is $C_e=4\pi
  M$.  Therefore $2\pi A_{AH}/C_e^2=1+\sqrt{1-j_{AH}^2}$, where
  $A_{AH}$ is the area of the apparent horizon.  Thus, the AH area and
  the equatorial circumference
  can be used to estimate the spin of the final BH from
  \cite{Kiuchi:2009jt}
  \be
  \label{eq:spinArea}
  j_{AH} = \sqrt{ 1 - \left(\frac{2\pi A_{AH}}{C_e^2} - 1 \right)^2 } \, .
 \ee
\end{enumerate}
%
%
\begin{table*}
\begin{tabular}{|c|c|c|c|c|c|c|c|}
\hline
Run  & Grid Setup   & $R_B/M$   & $d/M$    & $M_{{\rm irr},i}/M$   & $P_{i}/M$  &
$J^{AH}_{\rm fin}/M^2$ & $J^{rad}_{\rm fin}/M^2$    
\\ \hline
IN1 & $\{(48,24,12,6)\times (1.5, 0.75),~h=1/56\}$ & $40$ & $6.517$ & $0.483$ & $\pm 0.133$ & $0.69$ & $0.70$  \\
\hline
IN2.1 & $\{(48,24,12,6)\times (1.5, 0.75),~h=1/48\}$ & $30$ & $6.517$ & $0.483$ & $\pm 0.133$ & $0.69$ & $0.65$  \\
IN2.2 & $\{(48,24,12,6)\times (1.5, 0.75),~h=1/52\}$ & $30$ & $6.517$ & $0.483$ & $\pm 0.133$ &  & $0.65$  \\
IN2.3 & $\{(48,24,12,6)\times (1.5, 0.75),~h=1/56\}$ & $30$ & $6.517$ & $0.483$ & $\pm 0.133$ &  & $0.65$  \\
\hline
HD1   & $\{(48,24,12,6)\times (1.5, 0.75),~h=1/60\}$ & $40$ & $6.517$ & $0.483$ & $0.0$ & $0.0$ & $0.0$ \\
\hline
VIS & $\{(48,24,12,6)\times (1.5, 0.75),~h=1/48\}$ & $48$ & $6.517$ & $0.483$ & $\pm 0.133$ & &  \\
\hline
\end{tabular}
\caption{\label{tab:sphereruns} Grid structure, as well as initial and final
parameters of the simulated black holes. The grid setup is given
in terms of the radii of the individual refinement levels as well as the
resolution near the punctures $h$ (see Sec.~II E in \cite{Sperhake:2006cy}
for details). The reflective outer boundary is located at radius
$R_B$. The table further shows the initial coordinate separation of the
two punctures $d$, the irreducible mass $M_{\rm irr}$ and the Bowen-York
\cite{Bowen:1980yu} parameter for initial linear momentum $P_i$
of the individual holes. $J^{AH}_{\rm fin}$ and $J^{rad}_{\rm fin}$
are the spin of the single hole after merger determined from the AH and
the merger radiation, respectively. All parameters are given in units of the 
ADM mass. We did not monitor the AH properties for all runs, therefore the spin of the
final BH is not determined (empty cells) for some cases.
The Weyl scalars have been extracted at $r_{\rm ex} = 35M$ (IN1, HD1) and
$r_{\rm ex} = 25M$ (IN2), respectively. Model VIS is used in
Appendix \ref{sec:appsnapshots} for visualisation.
}
\end{table*}
%

\section{Numerical results\label{sec:num}}

Our numerical study focuses on two types of binary BH
initial configurations; (i) head-on collisions of nonspinning
BHs starting from rest and (ii) quasicircular inspiral of
nonspinning holes.
In the remainder of this work we label these as HD and IN simulations.
The initial parameters of all our simulations
as well as the structure of the computational domain and the position of the
outer boundary $R_B$ are summarised in Table \ref{tab:sphereruns}.
Unless denoted otherwise, the results presented refer to the highest
resolution available.

\subsection{\label{sec:convergence}Numerical convergence analysis}
%
\begin{figure*}
\begin{center}
\begin{tabular}{cc}
\includegraphics[clip=true,width=0.45\textwidth]{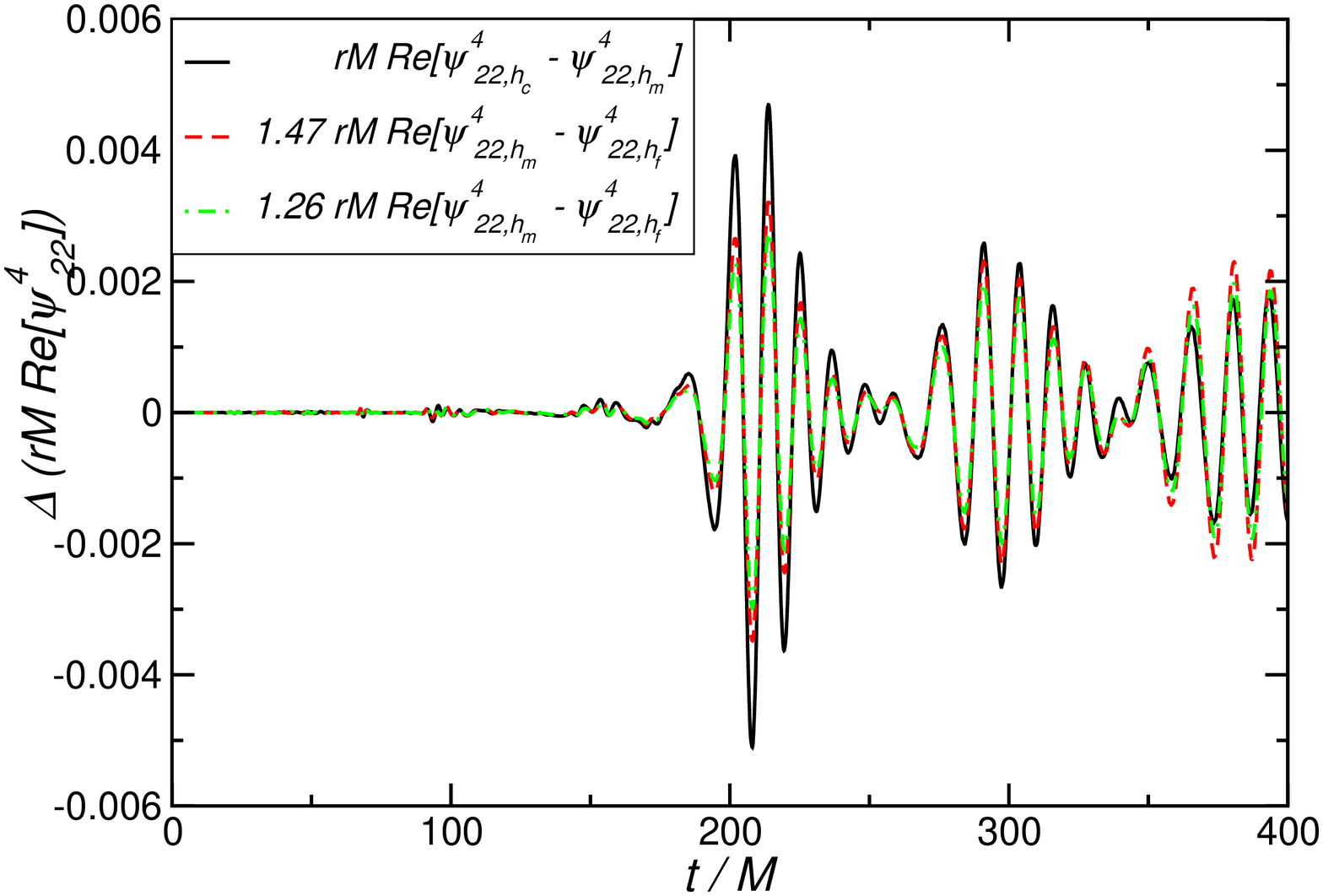} &
\includegraphics[clip=true,width=0.45\textwidth]{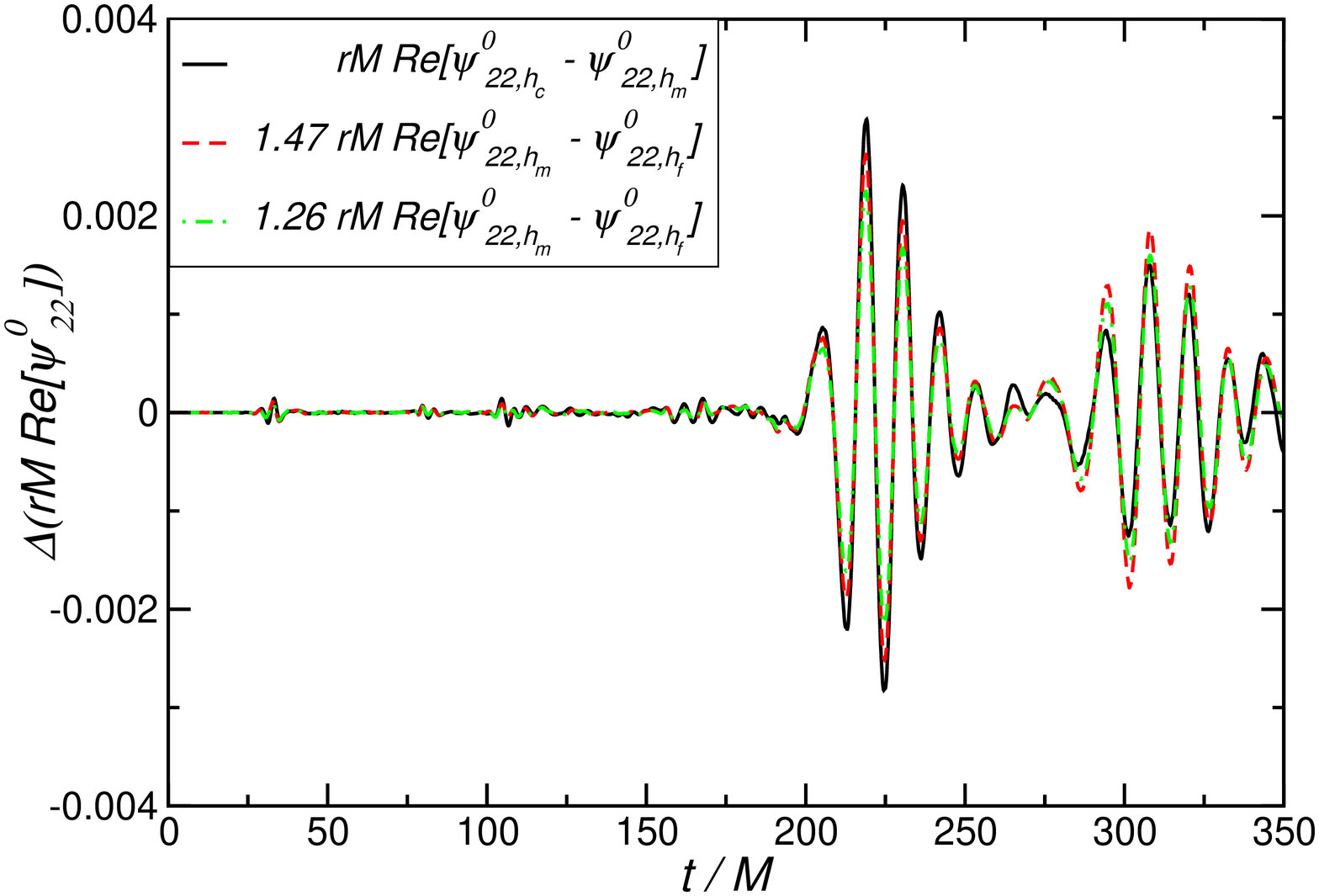}
\end{tabular}
\end{center}
\caption{\label{fig:convergencePsi} Convergence analysis of the outgoing
Weyl scalar $\Psi_4$ (left panel) and the ingoing Weyl scalar $\Psi_0$
(right panel) for the IN2 runs.  We show the differences of the
$l=m=2$ mode between the coarse and medium and the medium and fine resolution
run. The latter has been amplified by the factors $Q=1.47$ (fourth-order
convergence) and $Q =1.26$  (second-order convergence).  We observe fourth-order 
convergence in the signal due to the merger whereas the first and
second after-merger cycles show only second-order convergence. The first
two reflected and ingoing wave pulses show second-order convergence.
}
\end{figure*}
Before we discuss in detail the physical properties of the BH
binary systems, we test the accuracy of our simulations by performing
a convergence analysis of model IN$2$.
Numerical simulations
based on the finite differencing method typically approximate
the continuum solution of differential equations with a
leading error
that has a polynomial dependence on the resolution, $f_{\rm cont}
= f_{\rm num} + \mathcal{O}(h^n)$. The order $n$ depends on the
specific numerical implementation. Consistency of the code
is tested by evolving the same configurations with low,
medium and high resolution $h_c$, $h_m$ and $h_f$. One straightforwardly
shows that the convergence factor is then given by
\begin{equation}
  Q\equiv \frac{f_{h_c} - f_{h_m}}{f_{h_m} - f_{h_f}}
      = \frac{h_c^n - h_m^n}{h_m^n - h_f^n},
\end{equation}
where $f_h$ stands for any of the evolved variables obtained for
resolution $h$.
For the case of contemporary moving puncture codes, the
solution is complicated by the fact that the differential equations
are typically discretized with fourth- (or higher) order accurate
stencils but prolongation in time between different refinement levels
and implementation of outer boundary conditions
is only second-order accurate; see Sec.~IV in Ref.~\cite{Brugmann:2008zz}
and our discussion in Sec.~\ref{sec:formulation}.

Fig.~\ref{fig:convergencePsi} shows our convergence results for the variables
$\Psi_4$ and $\Psi_0$ obtained for resolutions $h_c=M/48$, $h_m=M/52$ and
$h_f=M/56$. Here the differences $f_{h_m} - f_{h_f}$ have been amplified by the
convergence factors $Q=1.26$ and $1.47$ expected for second and fourth-order
convergence, respectively. The figure indicates fourth-order accuracy for the
first passage of the wave pulse and a gradual deterioration of convergence to
second-order accuracy afterwards. We believe this to be a consequence of the
different ingredients of the code as discussed above. At early stages, the
dominant error is the discretization of derivatives. As the pulse successively
passes across mesh refinement boundaries and is reflected off the outer
boundary, however, the second-order error in the
prolongation operation becomes dominant and reduces the order of convergence.
We also note, in this context, that well-posedness of the BSSN evolution system
with reflective boundary condition has so far not been demonstrated\footnote{To
  our knowledge, the well-posedness of the system of equations in combination
  with reflecting boundary conditions, as treated here, has not been studied
  yet. Some investigations of the wave equation with this type of boundary
  conditions suggest that it may be ill-posed \cite{Gustafsson1995,David}. These
  investigations also show that the wave equation with periodic boundary
  conditions is a well-posed initial boundary value problem
  \cite{Gustafsson1995,David}, pointing towards interesting future extension of
  our work.}. We can therefore not rule out adverse effects on the long-term
convergence properties due to potential ill-posedness of the continuum system of
equations. In the remainder of this discussion we will restrict ourselves to 2-3
passages of the wave pulse as covered in Fig.~\ref{fig:convergencePsi} during
which the relative uncertainties in $\Psi_4$ and $\Psi_0$ are $\le 5\%$.

\subsection{\label{ssec:waveforms}Gravitational wave signal and black hole
dynamics}
To our knowledge, this work presents the first analysis of gravitational
waveforms with {\em both} outgoing ($\Psi_4$) and ingoing ($\Psi_0$)
contributions for long-term stable numerical simulations
of BH binaries. For this reason, we first illustrate the
general pattern of the wave signal obtained for model VIS of
Table \ref{tab:sphereruns}. A series of snapshots of both
Newman-Penrose scalars are shown in Fig.~\ref{fig:snapshots}
in Appendix~\ref{sec:appsnapshots}
in superposed form.
\begin{figure}[tb]
\includegraphics[clip=true,width=8cm]{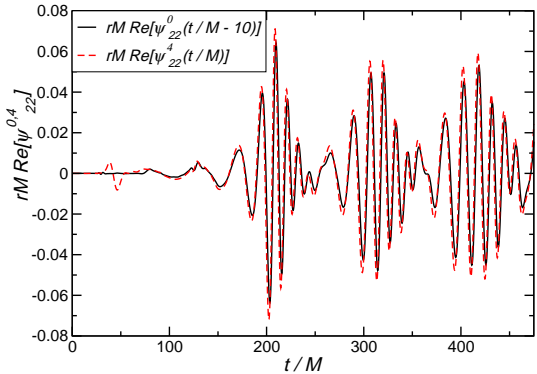}
\caption{\label{fig:comparePsi0Psi4_l2m2} Real part of the $l=m=2$ mode of
$rM\Psi_0$ and $rM\Psi_4$ of run IN1. The ingoing signal
$rM\Psi_0$ has been shifted in time by $\Delta t = 10M$ and in phase by $\pi$
(thus equivalent to an extra minus sign)
to account for the additional propagation time and the reflection.
}
\end{figure}

The gravitational wave signal is dominated by the quadrupole contributions
which we show in Fig.~\ref{fig:comparePsi0Psi4_l2m2}. For clarity, the
ingoing signal $\psi^0_{22}$ has been shifted in time by $\Delta t=10~M$
in order to compensate for the additional propagation time from the
extraction radius $r_{\rm ex}=35~M$ to the boundary $R_B=40~M$ and back
after reflection. The reflection introduces an additional phase shift
of $\Delta \phi = \pi$ which has also been taken into account in the
figure. Within numerical errors, we find
the resulting outgoing and subsequent ingoing pulses to overlap.

The first outgoing wave pulse, visible in Fig.~\ref{fig:comparePsi0Psi4_l2m2} around $150\le
t/M \le 250$, is generated during the inspiral, plunge and merger of
the binary and is similar to waveforms obtained for the inspiral of
nonspinning BH binaries in asymptotically flat space-times
(cf.~Fig.~1 in Refs.~\cite{Sperhake:2006cy,Brugmann:2008zz}).
Due to the reflecting boundary, however, this wave pulse does not escape
the computational domain. Instead it propagates inwards, interacts
with the post-merger remnant hole and eventually manifests itself as
a second wave pulse shifted by $\Delta t\approx 80~M$ relative to the
first. This process repeats itself many times, with the wave pulse being
presumably distorted (by absorption, superradiance and other curved space-time effects on wave propagation) upon each interaction with the BH. We now investigate in detail these changes of the wave pulse upon interaction with the BH.
\subsubsection{Interaction of the wave pulse with the remnant black hole}
\begin{figure*}
\begin{center}
\begin{tabular}{ccc}
\includegraphics[clip=true,width=0.3\textwidth]{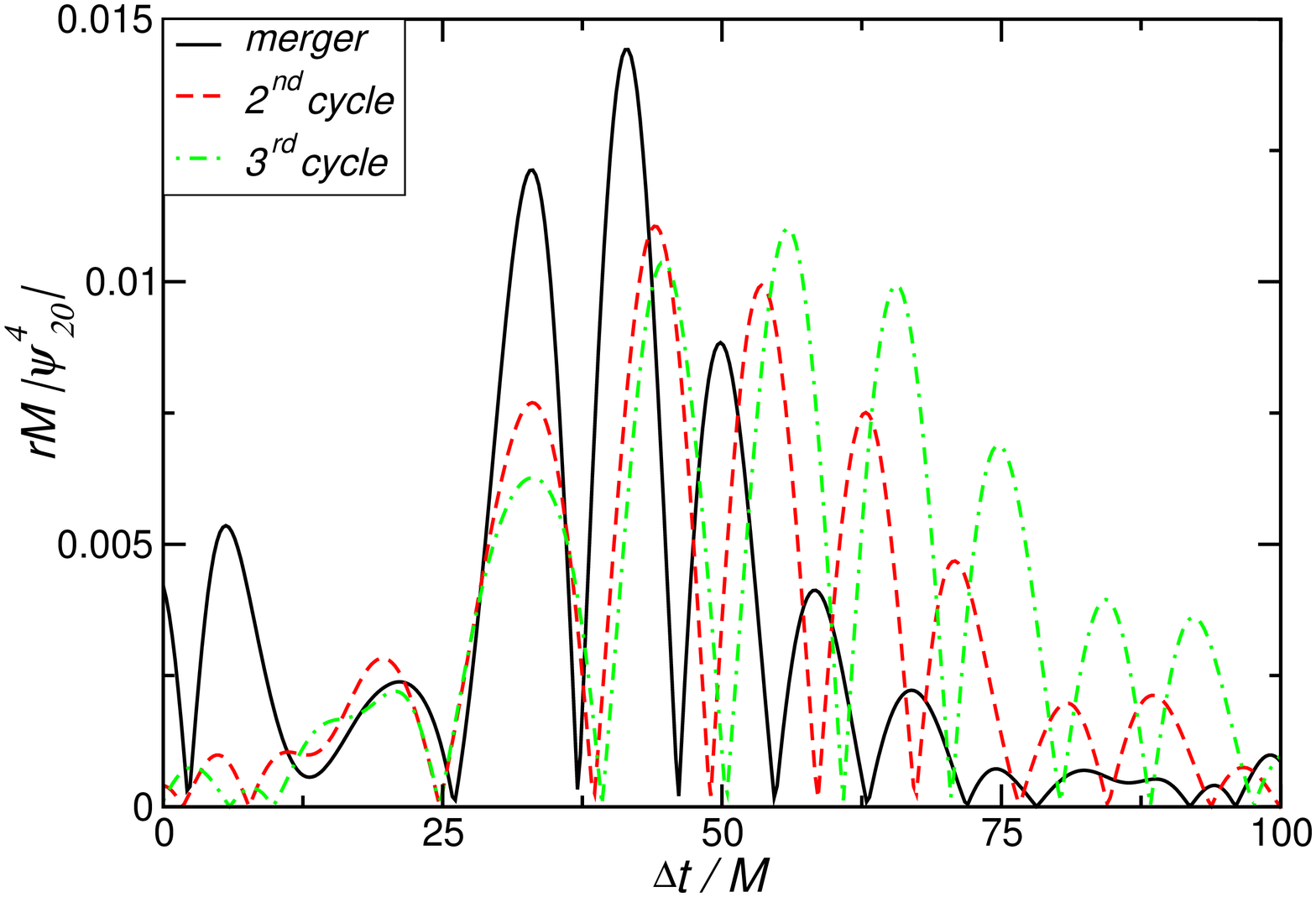} &
\includegraphics[clip=true,width=0.3\textwidth]{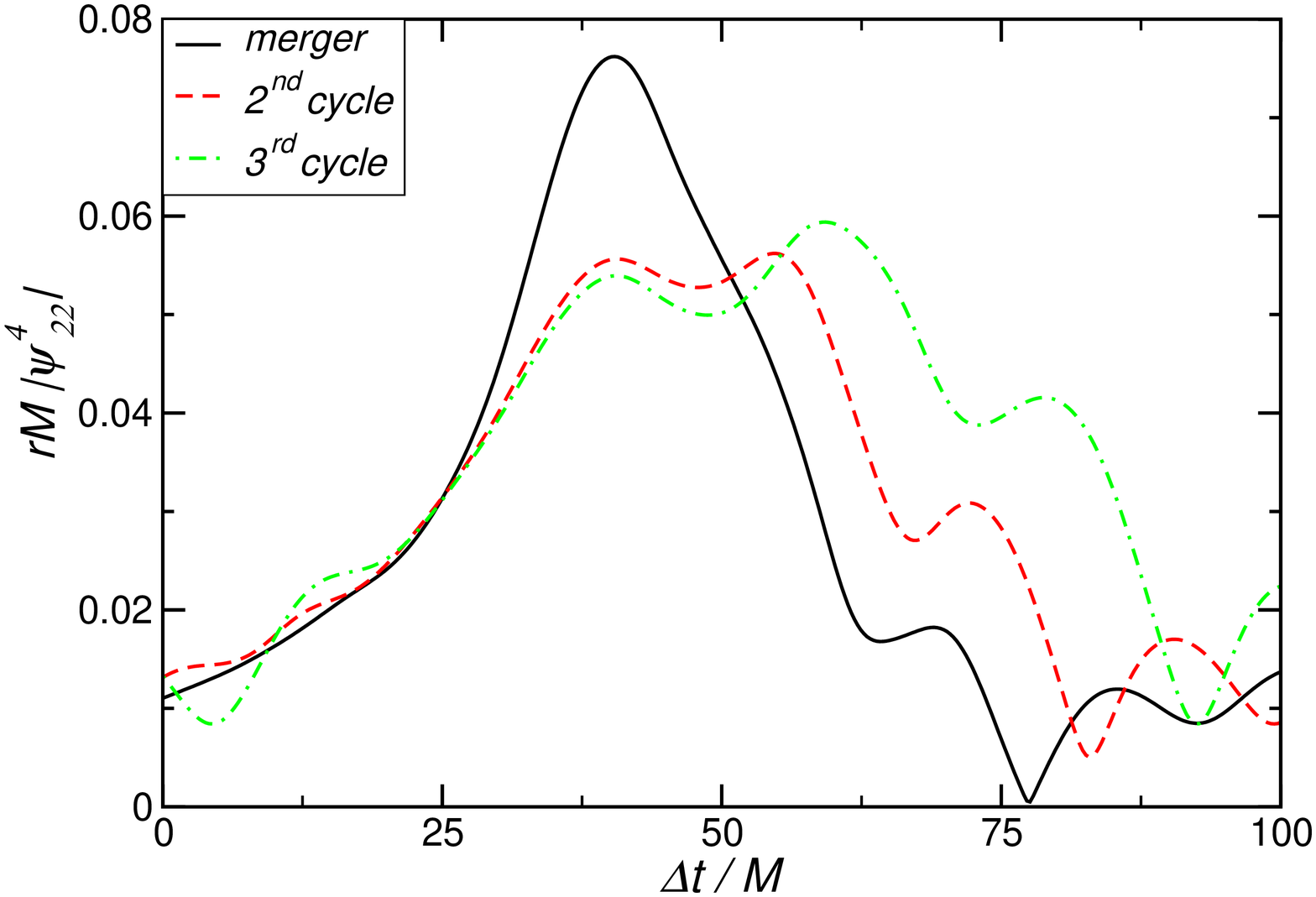} & 
\includegraphics[clip=true,width=0.3\textwidth]{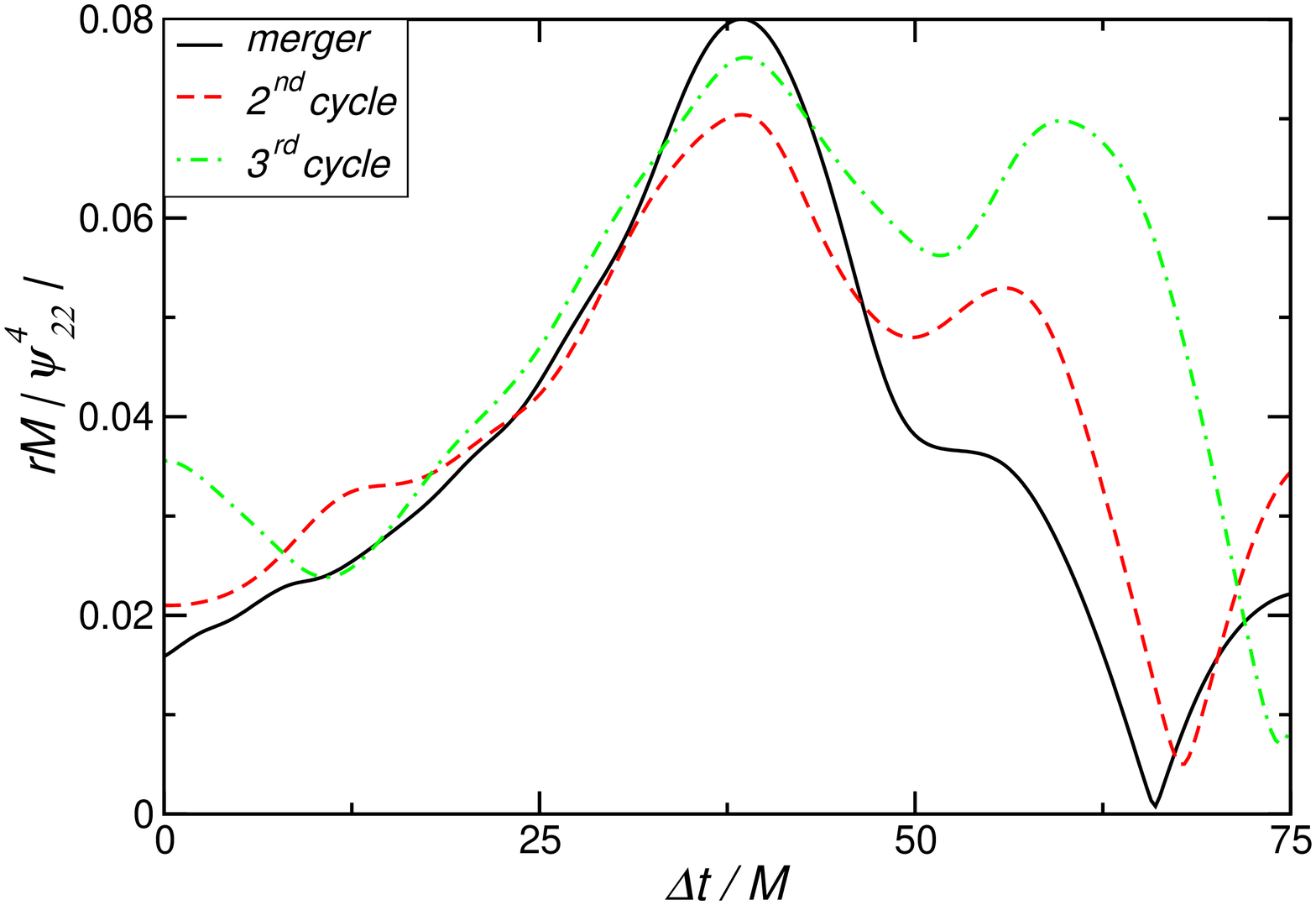} 
\end{tabular}
\end{center}
\caption{\label{fig:overlapshifted} Overlap of the amplitudes of
successive pulses of the same waveform; $l=2$,$m=0$ for the HD1 run (left),
$l=m=2$ for the IN1 (centre) and IN2.3 (right panel) simulations,
obtained by time-shifting such that the maxima overlap.}
\label{fig:overlap}
\end{figure*}

As shown in Fig.~\ref{fig:comparePsi0Psi4_l2m2}, the outgoing
and subsequent ingoing wave pulses overlap within numerical uncertainties.
We therefore focus on the outgoing signal
in our study of subsequent wave pulses and the gradual changes
caused by successive scattering off the BH.
Changes in the wave pulse are best illustrated by considering the wave
amplitude as shown in Fig.~\ref{fig:overlapshifted}. Here we superpose
the $l=2$, $m=0$ mode for model HD1 and
the $l=2$, $m=2$ multipoles for models IN1 and IN2
of the first three successive outgoing
wave pulses by applying corresponding time shifts to the waveform.
Clearly, the wave pulses broaden after each scattering off the BH.
We emphasise that this distortion of the pulse is {\em not} an artifact
of the outer boundary condition as is demonstrated by the good overlap
between the ingoing and outgoing pulses in Fig.~\ref{fig:comparePsi0Psi4_l2m2}.
\begin{figure}
\includegraphics[width=8cm,clip=true]{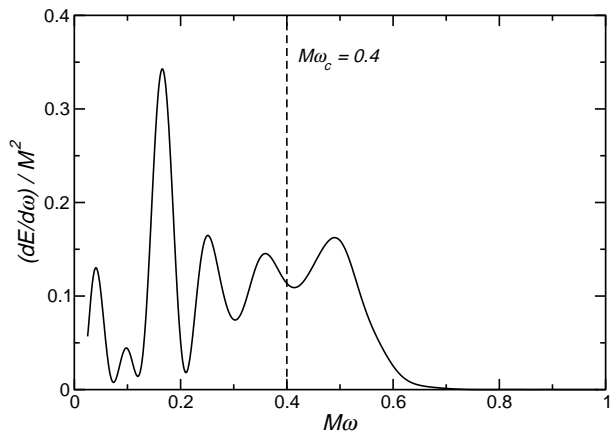}
\caption{\label{fig:spectrum} The energy spectrum for the $l=m=2$
multipole of the outgoing scalar $\Psi_4$, for model IN2.3.
This spectrum corresponds solely to the initial wave packet, i.e.,
the waveform has been truncated immediately before the first reflection
off the boundary.  The vertical line marks the threshold frequency for
superradiance.
}
\end{figure}

One possible explanation for this distortion relies on strong field,
curved space-time effects: massless waves in curved geometries do not
propagate {\it on} the light-cone.  Indeed, as shown in the classical
work by DeWitt and Brehme, the Green's function for a massless field
in a curved space-time does not generally vanish inside the light cone
\cite{DeWitt:1960fc}. This gives rise to interesting effects.  It is
well-known that wave propagation in BH space-times shows that the signal
can roughly be divided in three parts: (i) the first part is the prompt
response, at very early times, whose form depends strongly on the initial
conditions. This is the most intuitive phase, being a counterpart of the
light cone propagation in flat space-time; (ii) at intermediate times
the signal is dominated by an exponentially decaying ringing phase,
and corresponds to the excitation of the BH's characteristic modes of
vibration \cite{Berti:2009kk}; (iii) a late-time tail, usually a power
law falloff of the field \cite{Ching:1995tj,Cardoso:2003jf}.  Therefore,
a variety of possible distortions are possible.

Another possible explanation for the increasing width of the pulse arises in the
context of the {\em superradiance mechanism}.  It is well-known that the
scattering of a wave pulse off a rotating Kerr BH will result in superradiant
scattering---amplification of the scattered wave packet via extraction of
rotational energy from the BH---if the wave pulse satisfies
\begin{equation}
  \omega<m\Omega \, , \label{eq:omthresh}
\end{equation}
where $\Omega \equiv j_{\rm fin}/(2r_+)$ is the BH angular velocity
\cite{Bardeen:1972fi,Teukolsky:1974yv,Cardoso:2005vk}. We note, however,
that Fig.~\ref{fig:overlap} also indicates a broadening
of scattered wave pulses in the head-on case where no superradiance is
expected. While the observed broadening would be compatible with
superradiance, other effects appear to also influence the shape of
the pulse and our observations do not conclusively demonstrate its presence.

In order to investigate this dispersion in more detail, we plot in
Fig.~\ref{fig:spectrum} the energy spectrum for the dominant
$l=2$, $m=2$ mode.
The vertical line in this figure denotes the threshold frequency
$m\Omega \approx 0.4/M$ corresponding to a final spin $j_{\rm fin} =0.69$
as obtained for the post-merger hole for configuration IN2;
cf.~Sec.~\ref{sec:AH}. The figure demonstrates that the $l=2$,
$m=2$ mode does contain contributions which would be subject to
superradiance-induced amplification. These results then suggest that
the low-frequency component of the pulse is amplified due to superradiance,
while the high-frequency component is absorbed. A linear
analysis of superradiance in the Kerr geometry \cite{Teukolsky:1974yv}
shows that superradiant effects are always small, unless the hole is
rotating close to the extremal value. Thus, further studies,
including larger spins of the post-merger hole, are necessary to
comprehensively demonstrate superradiant wave amplification.

\subsubsection{\label{ssec:ahdata}Black hole dynamics}
%
\begin{figure}
\includegraphics[width=8cm,clip=true]{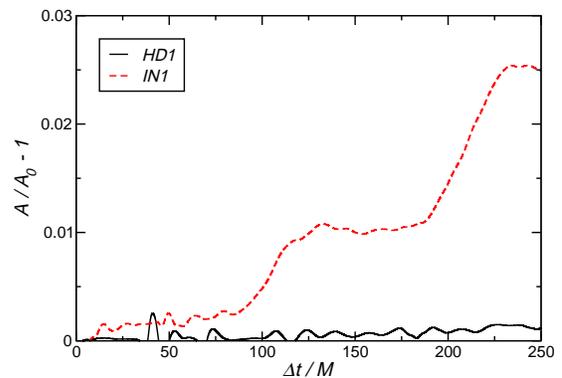}
\caption{\label{fig:AHarea} Time evolution of the area of the apparent
horizon for the head-on and inspiral simulation HD1 (solid curve) and IN1 (dashed curve).
The area of the rotating BH increases at regular intervals
corresponding to the propagation time of the pulse between the hole and
the reflective boundary at $r_{\rm ex}=35~M$.
Due to the small amount of radiation
generated during the plunge 
in the head-on case (HD1), the variation in the AH area
is buried in numerical noise.
}
\end{figure}
In contrast to the case of asymptotically flat space-times, a BH
binary under the influence of a reflective shell
does not settle down into a stationary configuration
soon after merger. This is due to the repeated interaction with the
wave pulse passing back and forth across the spatially
finite space-time. The
prolonged dynamical state of the system manifests itself prominently in
the area $A_{\rm AH}$ of the apparent horizon. In Fig.~\ref{fig:AHarea}
we show the fractional deviation $(A_{\rm AH}-A_0)/A_0$ of the apparent
horizon area from its value $A_0$ immediately after merger, i.e. the
first instance a common apparent horizon is found.
As expected, the horizon area remains nearly constant for the duration
of the first passage of the pulse to the reflective shell and back,
$\Delta t \approx 80~M$ and $60~M$, respectively, for models IN1 and
IN2.1\footnote{Small oscillations in
the horizon area are due to numerical uncertainties.}.
The subsequent increase in $A_{\rm AH}$ demonstrates that some fraction of the
gravitational wave pulse energy is absorbed by the hole. It follows another
period of approximately constant horizon area, a further increase
upon the second scattering of the pulse and so on.
Unfortunately, the radiation efficiency is almost 2 orders of magnitude
lower for
head-on collisions (see Table II in Ref.~\cite{Sperhake:2006cy}),
so that the increase in horizon area is buried in the numerical uncertainties.
The head-on case serves as a useful comparison, however, as it demonstrates
that the changes observed for the inspiralling configurations are significant
relative to numerical uncertainties.

\begin{figure}
\includegraphics[width=8cm]{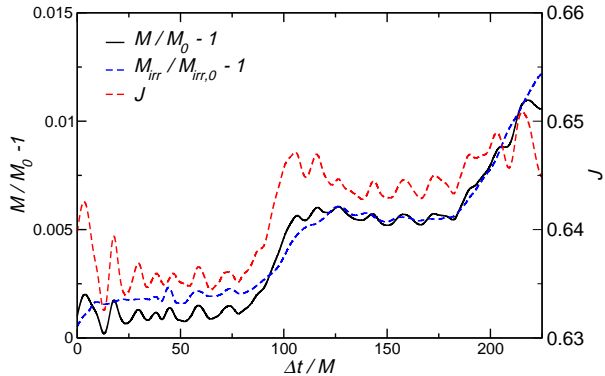}
\caption{\label{fig:AHmass}
Time evolution of the (relative) mass of the BH (solid line) computed by $M =
C_e/4\pi$, the irreducible mass (dashed line) and the total
spin $J = j M^2$ (dashed-dotted line).
}
\end{figure}

The BH mass, defined in terms of the
equatorial radius of the horizon $C_e$ by \cite{Kiuchi:2009jt}
\begin{equation}
M=\frac{C_e}{4\pi}\,,
\end{equation}
shows a similar behaviour as the horizon area. In Fig.~\ref{fig:AHmass}
we plot the fractional deviation $(M-M_0)/M_0$ of the mass
from its value immediately after merger together with the irreducible
mass and the BH spin $J$ of the hole obtained for model IN1.
The mass remains approximately constant until the pulse returns after
its first reflection, then increases, remains constant during the second
passage of the pulse and so on. In contrast, the spin shows
a significant increase only during the first scattering of the pulse off the BH.

We conclude that in each interaction with the gravitational radiation, the
final BH mass increases.  It is interesting to compare the increase in the
horizon mass with the amount of gravitational wave energy radiated during
the last stages of the inspiral, plunge and merger of a corresponding
binary system in an asymptotically flat space-time which is about $3.5~\%$
of the total energy of the system \cite{Berti:2007fi,Sperhake:2006cy}.
For the IN1 run, we estimate that about $15\%$ of the energy emitted
during the merger is absorbed by the central spinning BH per interaction.
Our results are consistent with total energy conservation. Moreover, they are {\it
not} incompatible with superradiant amplification:
typically, absorption of high-frequency waves is more effective than
superradiant amplification of low-frequency waves, such that the
net effect leads typically to absorption by the BH
\cite{Teukolsky:1974yv}.
The prolonged nonstationary character of the post-merger state is also
demonstrated by the time dependence of the BH's final spin.
Immediately after merger, we obtain $j_{\rm fin}= 0.69$ from
Eq.~(\ref{eq:spinArea}) in excellent
agreement with corresponding simulations using outgoing
radiation boundary conditions \cite{Berti:2007fi}. Successive interaction
with the reflected wave pulse, however, results in a small but significant
increase in the BH's spin as shown in Fig.~\ref{fig:AHmass}.
As before, the first increase occurs about $\Delta t=80~M$ after merger,
when the pulse has returned to the BH. We estimate
the fractional increase in spin resulting from the
first scattering at about $5~\%$. For comparison, the total
angular momentum radiated in the case of an asymptotically flat space-time
is reported as $J^{\rm rad}/M^2=0.246$ ($28~\%$ of the initial orbital angular momentum of the system) 
in Table I of Ref.~\cite{Berti:2007fi}.
It thus appears that a significant amount of angular momentum
remains in the form of gravitational waves.
Due to numerical uncertainties
it is not entirely clear whether later periods of interaction between
pulse and hole result in a further transfer of angular momentum
from the wave pulse to the hole or \textit{vice-versa}. Our results indicate,
however, that the amount of angular momentum exchanged in subsequent
interactions is significantly below $5~\%$.
\section{Discussion\label{sec:conclusions}}
The dynamics of BHs in generic space-times is a fascinating, yet extremely
challenging problem. The gauge/gravity duality, however, strongly
motivates us to solve dynamical problems with BHs on asymptotically AdS
backgrounds. In this paper, we have studied a toy model that captures
one of the fundamental features of such backgrounds: the active role
played by the boundary conditions for the bulk evolution.

We have mimicked the global structure of an AdS background by
introducing a reflecting wall at some radius. Within this cavity we
evolved an inspiralling BH binary and a BH binary starting from rest at
a certain initial distance. Of course, these are very specific initial
configurations, and serve merely as tests for future, possibly more
complex, situations.

Perhaps the most important conclusion of the present work is that these
simulations {\it can} be done and represent the first step to a full numerical
evolution of BHs in AdS space-times. Indeed, as observed in Section
\ref{sec:convergence}, it is not known whether the BSSN evolution scheme
together with reflecting boundary conditions is a well-posed initial value
problem. Thus, the convergence we have exhibited, which holds up to two reflections off
the boundary of the gravitational radiation produced in the merger, is the first
of our results.
Among the other results presented here we stress the following:
\begin{enumerate}
  \item
  For the first time, we were able to numerically study
  the scalar $\Psi_0$, describing ingoing waves, and check in the numerical
  data  the simple relations
  between $\Psi_0$ and $\Psi_4$ (cf. Eqs.~(\ref{eq:Psi4_ddh_defn}),
  (\ref{eq:Psi4_ddh_defn2}), (\ref{asymptotics}) and
  Fig.~\ref{fig:comparePsi0Psi4_l2m2}).

\item
  Our results are consistent with the intuitive
  expectations for a wave packet of radiation (generated during
  inspiral plus merger) travelling back and forth between the mirrorlike
  wall and the BH: part of this radiation is absorbed when
  interacting with the BH (especially high-frequencies). We estimate
  that about $15\%$ of the wave packet's energy is absorbed by the
  BH per interaction, at least during the first cycles.

  \item
    The wave packet is clearly distorted upon interaction
  with the hole, which means some frequencies are absorbed
  more efficiently than others. In principle, we should observe a small but nonzero
  superradiance. Unfortunately, we have not obtained incontrovertible proof
  of nonlinear superradiance from our numerical data. Presumably, the
  system will become unstable after a sufficiently long time, since the radiation should
  be exponentially amplified, once the high-frequency components had
  time to be completely absorbed by the BH. In future work
  we plan to investigate these instability studies further by considering
  a highly spinning, final BH produced by the inspiral of spinning
  BHs.

\end{enumerate}

One issue that we have not explored in this paper is the potential influence of
the box on the premerger dynamics. For sufficiently small size boxes, it is
plausible that the radiation produced in the inspiral may be reflected off the
boundary and interact with the binary \textit{before} the merger. This might
produce observable signatures in the premerger dynamics and even in the
properties of the remnant black hole. Whereas we seem to observe some hints of
this effect in our numerical data, a more exhaustive analysis is required to
produce some precise statements.

A future interesting extension of the present work is to repeat our analysis for
periodic boundary conditions.  Indeed, the wave equation in combination with
periodic boundary conditions is a well-posed initial boundary value problem
\cite{Gustafsson1995,David}. Such system is thus more likely to yield longer
stable and convergent numerical evolutions, allowing one to address a number of
interesting effects. Some relevant questions are: for how long is the numerical
evolution stable? Do the numerical instabilities set in before the system has
had time to reach a stable configuration? What are the typical times needed to
achieve this?

Also in the context of numerical relativity in space-times with a cosmological constant,
we plan on investigating dynamical black hole space-times in de Sitter backgrounds,
where many of the problems present in AdS do not exist 
(see also Refs.\cite{Beyer:2007vp,Beyer:2008gr,Beyer:2008gq} for some work along these lines).

Finally, it is necessary to extend this work in the obvious fashion: implement the evolution of BHs in real
AdS backgrounds. We hope the present work will help on achieving that.
\begin{acknowledgments}
%
     We thank Leonardo Gualtieri, David Hilditch and Florian Beyer
     for useful suggestions and discussions. We also thank the
     participants of the V Iberian Cosmology Meeting,
     the XII Marcel Grossmann Meetings, the Spanish Relativity
     Meeting 2009 and the I and II BH Workshop for useful feedback.
     M.Z. and H.W. are funded by FCT - Portugal through grants
     SFRH/BD/43558/2008 and SFRH/BD/46061/2008. A.N. is funded by FCT
     through grant SFRH/BPD/47955/2008. This work was supported by the
     {\it DyBHo--256667} ERC Starting Grant, by Funda\c c\~ao Calouste
     Gulbenkian, by FCT - Portugal through projects CERN/FP/109306/2009,
     CERN/FP/109290/2009, PTDC/FIS/64175/2006, PTDC/FIS/098025/2008,
     PTDC/FIS/098032/2008, PTDC/FIS/098962/2008, PTDC/CTE-AST/098034/2008
     by the Ram\'on y Cajal Programme of the Ministry of Education
     and Science of Spain, NSF grant PHY-0900735 and the Fairchild
     foundation to Caltech. This research was supported by an allocation
     through the TeraGrid Advanced Support Program under grant
     PHY-090003 and an allocation by the Centro de Supercomputaci{\'o}n
     de Galicia (CESGA) under project ICTS-2009-40. Computations were
     performed on the TeraGrid clusters TACC Ranger and NICS Kraken,
     at Magerit in Madrid, Finis Terrae, the Milipeia cluster in Coimbra,
     the Woodhen cluster at Princeton University and HLRB-II Garching.
     The authors thankfully acknowledge the computer resources, technical
     expertise and assistance provided by the Barcelona Supercomputing
     Centre---Centro Nacional de Supercomputaci\'on.
\end{acknowledgments}

\appendix

\section{Electromagnetic decomposition of the Weyl tensor}\label{sec:appWaveExtraction}
Since in this work we analyse the Newman-Penrose scalar $\Psi_0$,
besides $\Psi_4$, which is uncommon in numerical works, we collect in
this appendix some useful results. Following the sign convention in
\cite{Friedrich:1996hq, Sperhake:2006cy} the Newman-Penrose scalars
$\Psi_0$ and $\Psi_4$ are computed by eqs.~(\ref{eq:Psi0_Weyl}) and (\ref{eq:Psi4_Weyl}).
The vectors $\vec{k}, \vec{\ell}, \vec{m}, \bar{\vec{m}}$ form a null-tetrad. 
Their inner products vanish except for 
\be
-\vec{k}\cdot \vec{\ell} = 1 = \vec{m}\cdot\bar{\vec{m}}\,.
\ee
In practice, the vectors of the null-tetrad are constructed from a Cartesian
orthonormal basis $(u,v,w)$ in the spatial hypersurface
and the timelike orthonormal vector $\hat{\vec{n}}$ according to 
\begin{equation}
\begin{aligned}
  k^{\alpha} & = \frac{1}{\sqrt{2}} ( \hat{n}^{\alpha} + u^{\alpha}
  ), \\
  \ell^{\alpha} & = \frac{1}{\sqrt{2}} ( \hat{n}^{\alpha} -
  u^{\alpha} ),\\
  m^{\alpha} & = \frac{1}{\sqrt{2}} ( v^{\alpha} + i w^{\alpha}
  ), \\
  \bar{m}^{\alpha} & = \frac{1}{\sqrt{2}} ( v^{\alpha} - i
  w^{\alpha} ).
\end{aligned}
  \label{eq:nulltetradCart}
\end{equation}
The orthonormal triad vectors are constructed via the Gram-Schmidt orthonormalization starting with
\beq
u^i & = & [x,y,z], \\
v^i & = & [xz, yz, -x^2-y^2], \\
w^i & = & \epsilon^i_{jk} u^j w^k,
\eeq
where $\epsilon^i_{jk}$ is the 3-dimensional Levi-Civita tensor.
Next, we decompose the Weyl tensor in terms of its electric and magnetic parts 
\cite{Friedrich:1996hq}
\begin{equation}
\label{eq:WeylDecomp}
\begin{split}
  C_{\alpha\beta\gamma\delta} & = 2\left(
    l_{\alpha[\gamma}E_{\delta]\beta} - l_{\beta[\gamma}E_{\delta]\alpha} \right. \\
  & \quad \left. -
    \hat{n}_{[\gamma}B_{\delta]\tau}\epsilon^{\tau}_{\alpha\beta} -
    \hat{n}_{[\alpha}B_{\beta]\tau}\epsilon^{\tau}_{\gamma\delta}
  \right),
\end{split}
\end{equation}
where $l_{\mu\nu} = \gamma_{\mu\nu} + \hat{n}_{\mu}\hat{n}_{\nu}$ and
$\epsilon_{\alpha\beta\gamma} = \epsilon_{\mu\nu\lambda\rho}\hat{n}^{\mu}
\perp^{\nu}_{\alpha}\perp^{\lambda}_{\beta}\perp^{\rho}_{\gamma}$.
The electric and magnetic part of the Weyl tensor are given by
\beq
E_{\alpha\beta} & = & C_{\mu\nu\lambda\rho}\perp^{\mu}_{\alpha}\hat{n}^{\nu}\perp^{\lambda}_{\beta}\hat{n}^{\rho},\\
B_{\alpha\beta} & = & ^{\ast}C_{\mu\nu\lambda\rho}\perp^{\mu}_{\alpha}\hat{n}^{\nu}\perp^{\lambda}_{\beta}\hat{n}^{\rho}.
\eeq
$\perp^{\mu}_{\nu}$ denotes the projection operator onto the hypersurface and ${^{\ast}}$ denotes the Hodge dual.
By using the Gauss-Codazzi equations we express the electromagnetic components
in terms of the ``3+1'' variables \cite{Sperhake:2006cy}
\beq
\label{eq:elecWeylcomp3+1}
E_{ij}&=& R_{ij} - \gamma^{kl}( K_{ij}K_{kl}-K_{ik}K_{jl} )\,,\\
\label{eq:magnWeylcomp3+1}
B_{ij}&=& \gamma_{ik}\epsilon^{klm}D_l K_{mj}\,.
\eeq
If we insert Eq. (\ref{eq:WeylDecomp}), 
the definition of the null-tetrad (\ref{eq:nulltetradCart}) and the expressions 
(\ref{eq:elecWeylcomp3+1}), (\ref{eq:magnWeylcomp3+1}) into the definition of 
the Newman-Penrose scalars (\ref{eq:Psi0_Weyl}),(\ref{eq:Psi4_Weyl}) we obtain
\begin{widetext}
  \begin{equation}
    \label{eq:elmagnPsi}
    \begin{aligned}
      \Psi_0 & = \frac{1}{2} [ E_{kl} ( v^{k}v^{l} - w^{k}w^{l} )
      + B_{kl} ( v^{k} w^{l} + v^{l} w^{k} ] 
      +\frac{i}{2} [ E_{kl} ( v^{k}w^{l} + v^{l}w^{k} )
      - B_{kl} ( v^{k}v^{l} - w^{k}w^{l} ) ], \\
      \Psi_4 & = \frac{1}{2} [ E_{kl} ( v^{k}v^{l} - w^{k}w^{l} )
      - B_{kl} ( v^{k}w^{l} + v^{l}w^{k} ) ] 
      -\frac{i}{2} [ E_{kl} ( v^{k}w^{l} + v^{l}w^{k} ) +
      B_{kl} ( v^{k}v^{l} - w^{k}w^{l} ) ] \,.
    \end{aligned}
  \end{equation}
\end{widetext}
In the numerical code we use these relations in order to calculate 
$\Psi_0$ and $\Psi_4$ on the entire Cartesian grid. Then, they are 
interpolated onto coordinate spheres of various extraction radii $r_{\rm ex}$.
The Newman-Penrose scalars $\Psi_0$ and $\Psi_4$ are decomposed into 
spin-weighted spherical harmonics $_s Y_{lm}$ according to
\begin{align*}
  \psi^0_{lm}(t) & = \int d\Omega\Psi_0(t,\theta,\phi
  {_2} Y^{\ast}_{lm} \theta,\phi) \\
   & = (-)^{m+2}\int d\Omega \Psi_0(t,\theta,\phi)
  {_{-2}} Y_{lm}(\theta,\phi), \\
  \psi^4_{lm}(t) & = \int d\Omega \Psi_4(t,\theta,\phi) {_{-2}}
  Y^{\ast}_{lm}(\theta,\phi)\,.
\end{align*}
In the first equation we have used the relation \cite{Goldberg:1966uu}
\be {_s} Y^{\ast}_{lm}=(-)^{m+s} {_{-s}} Y_{lm}\,.
\ee
Thus, in practice we implement $\psi^0_{lm}$ and $\psi^4_{lm}$ only 
in terms of the spherical harmonics ${_{-2}} Y_{lm}$ with spin-weight $-2$.
\section{Snapshots}\label{sec:appsnapshots}
\begin{figure*}[p]
\begin{center}
\includegraphics[width=0.9\textwidth]{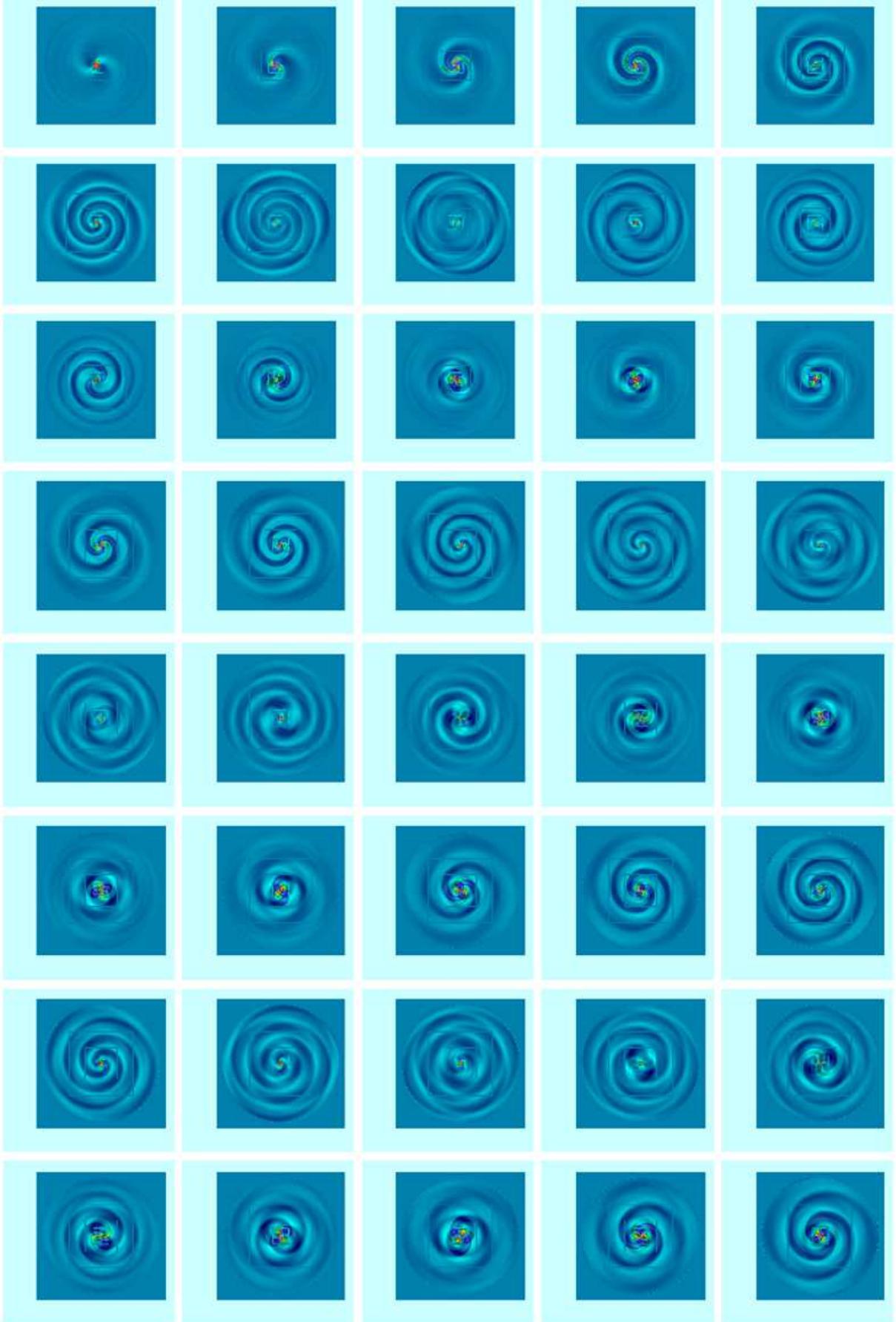} 
\end{center}
\caption{\label{fig:snapshots} Snapshots of $\Re(\Psi_4)$ superposed
by $\Re(\Psi_0)$.
The snapshots show the evolution from  $t=150M$ until $t=540M$ and
have a time interval of $\Delta t = 10M$. We show a slice of the orbital
plane with both coordinates going from $-48M,...,48M$. 
}
\end{figure*}

In Fig.~\ref{fig:snapshots} we illustrate the emission of the gravitational
wave signal during the inspiral and merger
and its evolution in the closed (confined) system containing a central,
spinning BH.
We display snapshots of the waveforms by superposing (the
real part of) $\Psi_0$ and $\Psi_4$ as obtained for
model VIS of Table \ref{tab:sphereruns}.
We show a slice of the orbital plane with $x, y = -48M,...,48M$
during an interval $t/M = 150, ..., 540$.
The difference in time between the individual pictures is $\Delta t/M = 10$.
The series of snapshots starts in the late inspiral phase shortly before
the plunge
and we see a strong gravitational wave signal that is emitted throughout
the merger (first row and first two columns of the second row).
This signal reaches the spherical boundary and is reflected back as
can be seen in the final three panels of the second row and first two
panels of the third row.
Starting with the third snapshot in the third row we see a second pulse
going outwards again after it has been scattered off the BH.
This process is repeated several times and the series of snapshots ends
with the
fourth outgoing wave pulse. An animation constructed from the numerical data can be found in~\cite{webpage}.
%
\section{Cubic-shaped box}\label{sec:appcubebox}

We have also performed simulations using condition (\ref{eq:refbc}) on a cubic
outer boundary.  These simulations represent the inspiral of nonspinning BH
binaries with initial separation $d/M = 6.514$ and initial linear momentum
$P_{y_i}/M = \pm 0.133$.  The grid setup for these runs is listed in Table
\ref{tab:cubicruns} together with a {\em reference} model O1 which describes the
inspiral of the same binary in an asymptotically flat space-time using outgoing
radiation boundary conditions.
\begin{table}[t!bh]
\begin{center}
\begin{tabular}{|c|c|c|c|c|}
\hline
Run  &  Grid Setup &  $h_f/M$  & $R_B/M$   & b.c. \\
\hline
O1 & (192, 96, 48, 24, 12, 6)(1.5, 0.75) & $1/40$ & $192$ & O \\
\hline
C1.1 & (24,12,6)(1.5, 0.75) & $1/40$ & $24$ & R \\
C1.2 & (24,12,6)(1.5, 0.75) & $1/44$ & $24$ & R \\
C1.3 & (24,12,6)(1.5, 0.75) & $1/48$ & $24$ & R \\
\hline
\end{tabular}
\caption{\label{tab:cubicruns} Parameters for a set of models evolved using a
  cubical boundary with reflective boundary condition (``R b.c.'') for models C1.1, C1.2 and
  C1.3 and an outgoing (``O b.c.'') Sommerfeld condition for model O1.  }
\end{center}
\end{table}
Gravitational waves have been extracted at $r_{\rm ex}=20\,M$
in the form of the Newman-Penrose scalar $\Psi_4$.

\begin{figure}[h!tpb]
\includegraphics[clip=true,width=8cm]{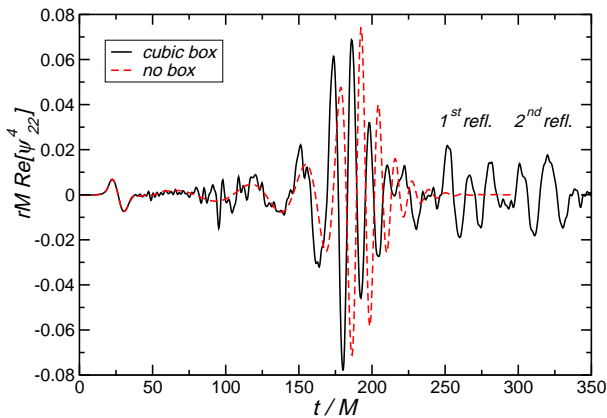}
\caption{\label{fig:Comp24cart} Comparison of the $l=2,m=2$ modes of
$\Psi_4$ obtained for models C1.1 (solid) and O1 (dashed curve).
The expected range in time for subsequent wave pulses
resulting from first and second reflections are indicated in the figure.
}
\end{figure}

In Fig.~\ref{fig:Comp24cart}, we compare the $l=m=2$ mode of $\Psi_4$ obtained
by the evolution of models C1.1 and O1.

\begin{figure}[h!tpb]
\includegraphics[clip=true,width=8cm]{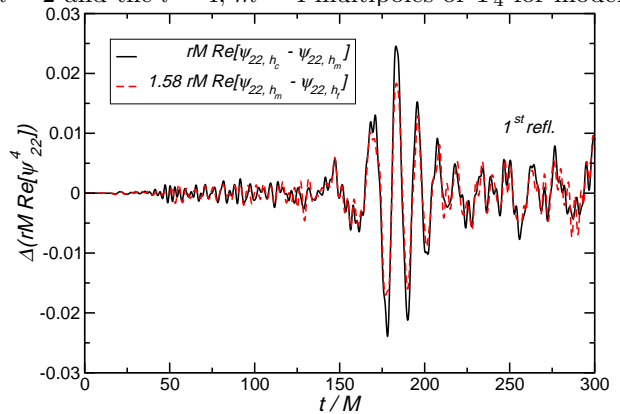}
\caption{\label{fig:convergence20cart} Convergence analysis of the $l=m=2$
mode of
$\Psi_4$ obtained for model {\it C1} of Table \ref{tab:cubicruns}.}
\end{figure}

\begin{figure*}[h!tbp]
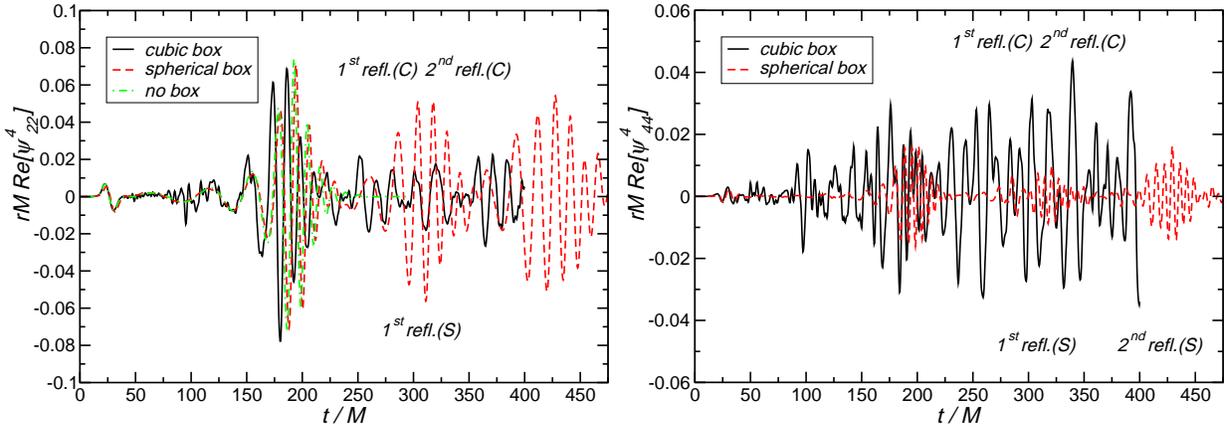

    \centering
    \subfloat{\includegraphics[width=0.45\textwidth]
      {fig15.eps}}
      \vspace{0.1cm}
    \subfloat{\includegraphics[width=0.45\textwidth]
      {fig16.eps}}
    \caption{\label{fig:ComparePsi4_lm} Comparison of the $l=m=2$
      (left) and $l=m=4$ (right) modes of $\Psi_4$ obtained
      for models C1.3 (cubic boundary), IN1 (spherical boundary)
      and O1 (outgoing condition).}
\end{figure*}

 In Fig.~\ref{fig:convergence20cart} we present
the convergence plot of the $l=m=2$ mode of $\Psi_4$ obtained from
evolutions of models C1.1, C1.2 and C1.3. The difference between the
medium and fine resolution result has been amplified by the factor
$Q=1.58$ corresponding to fourth-order convergence. While the
overall convergence is about fourth-order as in the case of a
spherical shell,
the cubical outer boundary
introduces a substantial amount of numerical noise which is also demonstrated
in Fig.~\ref{fig:ComparePsi4_lm} which displays the $l=2$, $m=2$ and
the $l=4$, $m=4$ multipoles of $\Psi_4$ for models C1.3 and IN1.
We believe that this is at least partly a consequence of mode mixing
in the case of the cubical boundary shell which is not well-suited for
the geometric shape of the gravitational wave pulse. For this reason,
we have exclusively used a spherical shell in the main part of this
work.


\bibliographystyle{h-physrev4}

\bibliography{num-rel}
  
\end{document}